\documentclass{article}
\usepackage{graphicx,subcaption} 
\usepackage{amsmath,amsfonts}
\usepackage{bbm}
\usepackage{comment,url}
\usepackage{float}
\usepackage{booktabs,multirow}
\usepackage[margin=1.5in]{geometry}

\RequirePackage{amsthm,amsmath,amsfonts,amssymb}
\RequirePackage[authoryear]{natbib}
\usepackage{authblk}
\usepackage{hyperref}

\title{Dyadic Models for Nonstationary Gene Flow in Landscape Genomics}

\author[1]{Michael R. Schwob}
\author[2]{Nicholas M. Calzada}
\author[3,4]{Justin J. Van Ee}
\author[5]{Diana Gamba}
\author[6]{Rebecca A. Nelson}
\author[7]{Megan L. Vahsen}
\author[8]{Peter B. Adler}
\author[9]{Jesse R. Lasky}
\author[4]{Mevin B. Hooten}

\affil[1]{Department of Statistics, Virginia Tech}
\affil[2]{Department of Statistics, The Ohio State University}
\affil[3]{Department of Public Health Sciences, University of Hawai`i at Mānoa}
\affil[4]{Department of Statistics and Data Science, The University of Texas at Austin}
\affil[5]{Department of Biology, University of New Mexico}
\affil[6]{Ecology, Evolution and Behavior Program, Michigan State University}
\affil[7]{Odum School of Ecology, University of Georgia}
\affil[8]{Department of Wildland Resources, Utah State University}
\affil[9]{Department of Biology, Pennsylvania State University}

\begin{document}

\maketitle

\begin{abstract}
    The field of landscape genomics aims to infer how landscape features affect gene flow throughout a population. Most landscape genomic frameworks assume the isolation-by-distance and isolation-by-resistance hypotheses, which propose that genetic dissimilarities between individuals increase as a function of distance and as a function of cumulative landscape resistance, respectively. While these hypotheses are valid in certain settings, other mechanisms may affect gene flow. For example, the gene flow of invasive species may depend on founder effects and multiple introductions. Such mechanisms are not considered in modern landscape genomic models. We extend dyadic models to allow for mechanisms that range-shifting and/or invasive species may experience by introducing dyadic spatially-varying coefficients (DSVCs) defined on pairs of sampled locations. The DSVCs allow the effects of landscape on gene flow to vary across space, capturing nonstationary and asymmetric connectivity. Additionally, we incorporate explicit landscape features as connectivity covariates, which are localized to specific regions of the spatial domain and may function as barriers or corridors to gene flow. Such covariates are central to colonization and invasion, where spread accelerates along corridors and slows across landscape barriers. The proposed framework accommodates colonization-specific processes while retaining the ability to assess landscape influences on gene flow. Our case study of the highly invasive cheatgrass (\textit{Bromus tectorum}) demonstrates the necessity of accounting for nonstationary gene flow in range-shifting species.
\end{abstract}


\section{Introduction}

Landscape genomic models aim to quantify the effects of environmental heterogeneity on neutral and adaptive genetic variation \citep{schwartz2010landscape}. Most models relate measures of genetic dissimilarity to environmental features to infer how the local environment impacts gene flow throughout a population \citep[e.g.,][]{hanks2013circuit,schwob2024composite}.
By linking genetic dissimilarity with spatial patterns of connectivity, landscape genomic models provide a statistical framework for inferring the processes that affect gene flow and population structure.
Such models are also used to identify environmental features that facilitate or impede movement \citep[i.e., corridors or barriers;][]{peterson2019spatially,white2020novel}, to test competing hypotheses about dispersal mechanisms \citep{aitken2024conserving}, and to predict the spread of populations across heterogeneous landscapes \citep{schwartz2010landscape}. While landscape genomic analyses are growing in popularity, most applications focus on native or threatened taxa; analyses on invasive species remain scarce \citep{briscoe2024isolation,kolodziejczyk2025genomic}. As a result, modern landscape genomic models often omit invasion processes that induce nonstationary gene flow (e.g., founder effects, multiple introductions, human-assisted transport). Accounting for such dynamics is critical for accurate inference of gene flow in range-shifting and/or invasive species. We present a landscape genomics framework specifically to infer the nonstationary gene flow of such species. 


A common approach to infer functional connectivity across space is the circuit-theoretic framework  \citep{mcrae2008using,hanks2013circuit}, 
in which the spatial domain is discretized into a circuit and represented as a network with resistance distances controlling connectivity between sample locations; resistance distance is often defined as a function of landscape and environmental features. Circuit theory has been widely applied to investigate gene flow and migration in a broad range of species \citep{hanks2016latent,fletcher2022extending,stack2026riverscape}. 
\cite{schwob2024composite} introduced a continuous-space analogue to circuit-theoretic methods by linking dyadic regression over a network with advection-diffusion differential equations to yield potential surfaces that characterize asymmetric gene flow. 
Accounting for asymmetry is particularly important in systems where directional processes shape movement and dispersal, such as for invasive species \citep{wilson2009something}.
We adopt the dyadic framework as the foundation for our model because it provides a principled approach for modeling asymmetric patterns of gene flow.

Most landscape genomic frameworks exclusively test two core hypotheses: isolation-by-distance (IBD) and isolation-by-resistance (IBR). IBD assumes that genetic dissimilarity increases with geographic distance \citep[][]{duforet2014nonstationary,shastry2025jointly}, and IBR assumes that the extent to which genetic dissimilarity increases depends on the effective resistance between locations \citep[i.e., the accumulated difficulty of traversing the landscape;][]{mcrae2006isolation}. Together, these two hypotheses have been used to infer how landscape features affect gene flow in many species \citep{mcrae2008using,novembre2008genes}.
However, invasive populations often violate the core assumptions in IBD and IBR. Multiple introductions \citep{vahsen2025phenological}, founder effects \citep{shirk2014patterns}, allele surfing \citep{goodsman2014genetic}, pulsed admixture \citep{chun2010gene}, and directional drift \citep[e.g., prevailing winds, rivers;][]{belouard2019genetic} are examples of invasion dynamics that yield nonstationary gene flow \citep{gamba2025local}. 
To model such heterogeneous and directional processes, we introduce dyadic spatially varying coefficients (DSVCs) that allow the effect of landscape features on gene flow to vary across space.

An additional source of nonstationarity is human-assisted transportation \citep{medley2015human}, which occurs along specific pathways (e.g., highways, railroads). 
Unlike landscape features that cover the entire spatial domain (e.g., elevation, precipitation), pathways are localized, not recorded at the node-level across the network, and can affect gene flow in ways that IBD and IBR cannot describe; these characteristics create sharp discontinuities in connectivity \citep{goel2025mechanistic} and extend to naturally occurring features, such as riparian corridors.
To represent such pathways, we introduce connectivity covariates, which we define as spatially-restricted predictors that attenuate connectivity along localized, discrete pathways within the spatial domain.
The proposed connectivity covariates can be seen as the dyadic analogue to the resistance surfaces used in circuit-theoretic frameworks.

In what follows, we introduce our landscape genomics case study in Section 2 and review dyadic modeling in Section 3. In Section 4, we extend dyadic models to account for invasion mechanisms by introducing DSVCs and connectivity covariates. We provide implementation details in Section 5 and a replicated simulation study in Section 6. In Section 7, we analyze cheatgrass (\textit{Bromus tectorum}) gene flow in its native and invaded ranges. Finally, we provide concluding remarks in Section 8.

\section{Cheatgrass Invasion in North America}

Cheatgrass is an invasive winter annual grass with a native range across Eurasia and northern Africa. Introduced to North America by the 1890s \citep{mack1981invasion}, it rapidly spread across arid and semi-arid ecosystems in the Intermountain West, where it is estimated to occur at high densities in 31\% of the region \citep{bradley2018cheatgrass}. Its invasion success stems from a short life cycle and rapid reproduction, displacing native perennials and degrading wildlife habitat \citep{porensky2016historical}, and it produces a highly flammable, continuous fuel bed that intensifies fire frequency and severity in landscapes that historically burned infrequently \citep{germino2016ecosystem}. 

Although the ecological impacts of cheatgrass are well-studied, much less is known about the evolutionary and demographic processes that have enabled its success in novel environments. To address this gap, we analyzed an extensive collection of cheatgrass genotypes across its native and invaded ranges. 
Previous analyses of these data have shown that pre-adapted genotypes facilitated invasion across North America \citep{gamba2025local} and that phenological sensitivity is influenced by current and source environments \citep{vahsen2025phenological}. Additionally, \cite{gamba2025local} found that landscape patterns of cheatgrass relatedness were consistent with IBD in the native range but not in the invaded range, suggesting the need to account for nonstationary gene flow in the invaded range.
However, important questions still remain about how population connectivity and spatial heterogeneity affect invasion dynamics.

We sought to understand how cheatgrass spread emerges from the combined influences of environment and connectivity. 
Because invasion is inherently relational, occurring between locations rather than within them, we adopted the dyadic perspective and modeled genetic dissimilarities between individuals as a function of pairwise environmental differences and spatial connectivity. Doing so let us infer how barriers, corridors, and ecological gradients give rise to asymmetric patterns of gene flow across heterogeneous landscapes. Additionally, we compared gene flow patterns across stable native populations with expanding invasive fronts to understand how nonstationary gene flow emerges from environmental differences and landscape features during invasion. 

\section{Dyadic Flow Models for Landscape Genomics}
\label{sec:dyadic-models}

We model gene flow by treating a sample of $n$ genomic sequences as a fully-connected network $\mathcal{N}=(\mathcal{V},\mathcal{E})$, where each genomic sequence is a spatially-referenced node in $\mathcal{V}=\{v_1,...,v_n\}$ and $\mathcal{E}=\{e_{12},...,e_{(n-1)n}\}$ comprises the $N={n \choose 2}$ edges that connect every pair of genomic sequences.
Each node has environmental covariates $\mathbf{x}_i\equiv\mathbf{x}(\mathbf{s}_i)\in\mathbb{R}^p$, where $\mathbf{s}_i\in\mathcal{S}$ denotes the location of node $i$ in geographic space and $\mathcal{S}\subset\mathbb{R}^2$ comprises the observed locations.
We let $\widetilde{y}_{ij}$ denote the weight of edge $e_{ij}$, which connects nodes $i$ and $j$; for a landscape genomics analysis, the edge weight $\widetilde{y}_{ij}$ is a measure of genetic dissimilarity between individuals $i$ and $j$.

We compute $\widetilde{y}_{ij}$ as the logit of the proportion of mismatched loci between nodes $i$ and $j$:
\begin{equation}\label{eq:pij}
    p_{ij}=\frac{d_{ij}+0.5}{M_{ij}+1},
\end{equation}
where $d_{ij}$ is the number of discordant loci between the genomic sequences and $M_{ij}$ denotes the number of loci comparable for that pair. 
The $+0.5$ continuity correction ensures finite logit transforms when $d_{ij}\in\{0,M_{ij}\}$, yields the bias-reduced estimate of \cite{firth1993bias} for separation in logistic regression \citep{heinze2002solution}, and corresponds to the Jeffreys prior posterior mean for a binomial likelihood \citep{chen2008properties}.
We compute genetic dissimilarity as $\widetilde{y}_{ij}=\text{logit}(p_{ij})$ because the logit-scale yields a response amenable to linear predictors and Gaussian processes.
The proposed framework is flexible to the choice of dissimilarity measure; we briefly discuss alternatives in Appendix A, including a binomial specification for the discordant counts $d_{ij}$ \citep{nei1972genetic,beaumont1998measuring}.

We use dyadic models to analyze the genetic dissimilarities $\{\widetilde{y}_{ij}\}$ because they allow us to study how landscape heterogeneity gives rise to asymmetric gene flow across space \citep{schwob2024composite}. Such models are similar to regressions of genetic and environmental distance matrices, which are common in landscape genomics \citep{balkenhol2009statistical,bradburd2013disentangling}. However, unlike conventional approaches, dyadic regression can be used to infer drivers of movement and identify barriers and corridors that govern gene flow throughout the landscape.

In dyadic models, each observation $\widetilde{y}_{ij}$ corresponds to an unordered pair of nodes $(i,j)$. 
We assume that $i<j$, which fixes an orientation for each edge.
For a fixed node $i$, the dyadic outcomes $\widetilde{y}_{ij}$ and $\widetilde{y}_{ik}$ are conditionally independent for $j\ne k$; dependence between dyads that share a node is accommodated through shared terms rather than through correlated residual error terms.
For example, we model the set of genetic dissimilarities $\{\widetilde{y}_{ij}=\text{logit}(p_{ij})\}$ as conditionally independent given pairwise differenced environmental covariates $\widetilde{\mathbf{x}}_{ij}=\mathbf{x}_j-\mathbf{x}_i$ and pairwise differenced latent effects $\widetilde{\eta}_{ij}=\eta(\mathbf{s}_j)-\eta(\mathbf{s}_i)$, where $\eta(\mathbf{s}_i)$ denotes the latent spatial random effect at the location of node $i$. Importantly, the pairwise differenced terms are signed, which allows us to recover directional flow despite using a response that is symmetric in $i$ and $j$.
We consider a conventional spatial dyadic regression for the $N={n \choose 2}$ unordered pairs in the index set$\{(i,j):i<j\}$:
\begin{equation}\label{eq:mini}
    \widetilde{y}_{ij}\sim \text{N}\left(\alpha+\mathbf{\widetilde{x}}_{ij}'\boldsymbol{\beta} + \widetilde{\eta}_{ij},\sigma^2\right),
\end{equation}
where $\alpha$ is the intercept (i.e., the baseline genetic dissimilarity shared across all dyads),
$\boldsymbol{\beta}\in\mathbb{R}^p$ are regression coefficients for the pairwise differenced environmental covariates $\widetilde{\mathbf{x}}_{ij}$, and $\sigma^2$ is the dyadic regression variance. 
The coefficients $\boldsymbol\beta$ quantify the marginal effect of observed environmental differences on genetic dissimilarity; because $\widetilde{\mathbf{x}}_{ij}$ is a signed difference, positive coefficients imply greater genetic dissimilarity when $x_j>x_i$, and negative coefficients imply the converse.

The pairwise differenced latent effects $\{\widetilde{\eta}_{ij}\}$ absorb residual spatial dependence not explained by the pairwise differenced environmental covariates, which ensures that inference on $\boldsymbol\beta$ is not confounded by residual spatial autocorrelation. We vectorize the node-level latent spatial effects $\boldsymbol{\eta}=(\eta(\mathbf{s}_1),\dots,\eta(\mathbf{s}_n))'$ and specify
$\boldsymbol{\eta}\sim \text{N}(\mathbf{0},\sigma_\eta^2\,\mathbf{R}(\phi_\eta))$, where $\sigma^2_\eta$ is the spatial variance that governs magnitude and $\mathbf{R}(\phi_\eta)$ is a spatial correlation matrix defined by the spatial range parameter $\phi_\eta$. Common choices for $\mathbf{R}(\phi_\eta)$ include the class of Matérn covariance functions, such as the exponential decay covariance function \citep{cressie2015statistics}.
Because $\text{Var}(\widetilde{\eta}_{ij})=2\sigma^2_\eta\{1-[\mathbf{R}(\phi_\eta)]_{ij}\}$, the expected magnitude of $\widetilde{\eta}_{ij}$ grows with the distance between $\mathbf{s}_i$ and $\mathbf{s}_j$ as in isolation-by-distance modeling, and $\phi_\eta$ governs the scale over which it grows.
Our approach to modeling latent spatial effects yields an induced dyadic covariance that is determined by $n$ node-level effects rather than by $N$ free parameters.

Although the responses $\{\widetilde{y}_{ij}\}$ are symmetric in $i$ and $j$, the mean structure in \eqref{eq:mini} is not. Reversing $(i,j)$ flips the sign of $\widetilde{\mathbf{x}}_{ij}$ and of $\widetilde{\eta}_{ij}$.
Therefore, the mean of a dyad separates into an even component that is unchanged under reversal ($\alpha$) and an odd component that changes sign ($\widetilde{\mathbf{x}}_{ij}'\boldsymbol{\beta}+\widetilde{\eta}_{ij}$).
Conventional landscape genomic models summarize a pair symmetrically because they have no odd component and, consequently, describe only the magnitude of gene flow. In contrast, the model in \eqref{eq:mini} has an odd component because the differenced covariates and latent effects are both signed; each pair is observed once and in one orientation.
Hereafter, we refer to the model in (\ref{eq:mini}) and its extensions as dyadic flow models; this directional perspective is critical for identifying sources and sinks, assessing unidirectional barriers (e.g., prevailing winds, rivers) and habitat fragmentation, and characterizing invasion dynamics where spread is inherently asymmetric. 

\cite{schwob2024composite} linked the model in (\ref{eq:mini}) with advection-diffusion differential equations such that the mean structure of the dyadic model can represent the gradient of a potential function. As a result, a posterior potential surface can be inferred, which describes the mechanistic flow of processes in space and time. 
In our application, dyadic flow models predict genetic dissimilarity and provide interpretable surfaces and flow fields that link observed genomic data to dispersal processes.
Variations of (\ref{eq:mini}) have also been used to study international relations \citep{minhas2022taking} and tuberculosis transmission \citep{warren2023spatial}, revealing the flexibility of dyadic flow models for studying directional processes.

\section{Accounting for Nonstationary Processes}
Because invasive species spread opportunistically through heterogeneous landscapes, models that characterize nonstationary processes are crucial for accurately representing invasion dynamics.
In what follows, we extend the dyadic flow model in (\ref{eq:mini}) for the study of invasive species by introducing dyadic spatially varying coefficients and connectivity covariates. DSVCs allow the influence of landscape features on gene flow to vary across space, characterizing nonstationary processes that are inherent to invasion dynamics. Connectivity covariates account for localized landscape features that create sharp discontinuities in landscape resistance, such as anthropogenic pathways that facilitate human-assisted transportation or riparian corridors that facilitate wildlife dispersal.

\subsection{Dyadic Spatially Varying Coefficients}
Spatially varying coefficients (SVCs) are coefficients modeled as spatial processes, allowing them to change smoothly across space rather than being fixed globally \citep{gelfand2003spatial}. As a result, SVCs account for processes that are inherently heterogeneous across space because they can accommodate nonstationary effects, local heterogeneity, directional dynamics, and environmental gradients that global models would miss \citep{kim2021generalized,doser2025modeling}. 

While SVCs allow for spatial heterogeneity at the level of single locations, many ecological and genomic processes that we aim to infer are dyadic in nature (e.g., gene flow, dispersal connectivity, movement-based similarity). As seen in (\ref{eq:mini}), the dyadic covariates $\widetilde{\mathbf{x}}_{ij}$ are pairwise differenced site-level covariates. Thus, any spatially varying coefficients for $\widetilde{\mathbf{x}}_{ij}$ must exist in the dyadic space $\mathcal{D}=\{(\mathbf{s}_i,\mathbf{s}_j)\mid \mathbf{s}_i,\mathbf{s}_j\in\mathcal{S},i<j\}$. We extend SVCs to the dyadic space by allowing regression coefficients to vary across pairs of locations:
\begin{equation}\label{eq:wSVCs}
    \widetilde{y}_{ij}\sim \text{N}\left(\alpha+\mathbf{\widetilde{x}}_{ij}'(\boldsymbol{\beta} + \boldsymbol\delta_{ij}) + \widetilde{\eta}_{ij},\sigma^2\right),
\end{equation}
where $\boldsymbol\beta\in\mathbb{R}^p$ are the fixed dyadic coefficients and $\boldsymbol\delta_{ij}\in\mathbb{R}^p$ are dyadic spatially varying coefficients (DSVCs) for the pairwise location set $(\mathbf{s}_i,\mathbf{s}_j)$. 

Naturally, one might consider a Gaussian process prior on the DSVCs; however, doing so would require ${n \choose 2}\times p$ parameters and becomes computationally infeasible as $n$ increases. 
To facilitate computation, we instead model the DSVCs using the latent spatial factor model
$\boldsymbol\Delta=\mathbf{WC}',$
where each row in $\boldsymbol\Delta\in\mathbb{R}^{N\times p}$ corresponds to $\boldsymbol\delta_{ij}'$, $\mathbf{W}\in\mathbb{R}^{N\times Q}$ are dyad-indexed latent spatial factors, and $\mathbf{C}\in\mathbb{R}^{p\times Q}$ are covariate loadings \citep{zhang2022spatial,doser2025modeling}. 
The latent spatial factor model addresses the dimensionality challenge by factorizing the DSVC surface. Instead of learning $N \times p$ coefficients, the spatial factor model allows us to learn a low-rank structure with $(N\times Q)+(p\times Q)$ parameters, where $Q\ll N$. 
We use a spatial factor model for the DSVCs because it achieves parsimony through dimension reduction while retaining enough flexibility to characterize nonstationary, covariate-specific patterns in dyadic space \citep{taylor2019spatial}.
For each latent spatial factor $\mathbf{w}_q$ ($q=1,\dots,Q$), we consider a Gaussian process prior in dyadic space:
$\mathbf{w}_q \sim \text{N}(\boldsymbol0,\boldsymbol{\Sigma}_q),$
where $\mathbf{w}_q$ is the $q$th column of $\mathbf{W}$ and $\boldsymbol{\Sigma}_q$ is a spatial covariance matrix defined by a kernel on $\mathcal{D}$. 
Thus, we model spatial structure directly in the dyadic space $\mathcal{D}$ rather than geographic space $\mathcal{S}$.

A dyadic covariance matrix $\boldsymbol\Sigma_q\in\mathbb{R}^{N \times N}$ defined on all $N={n \choose 2}$ dyads would be computationally infeasible for most $n$.
To facilitate computation, we assume a separable covariance structure in the full node-node space, then restrict the covariance to $\mathcal{D}$ with a single node-level kernel $\mathbf{K}(\phi_q)\in\mathbb{R}^{n\times n}$ for both ends of a dyad, where $\phi_q$ is the spatial range parameter in node-space.
A symmetric kernel is appropriate when the same spatial processes govern both ends of a dyad, as in most ecological and invasion settings, and it reduces the number of spatial range parameters needed to infer the spatial processes. We discuss the separable construction in Appendix B, which yields the following covariance that expresses how similar dyads $(i,j)$ and $(i',j')$ are under the latent spatial process:
\begin{equation}\label{eq:dyad-cov}
    [\Sigma_q]_{(i,j),(i',j')}=K^q_{ii'}K_{jj'}^q+K^q_{ij'}K^q_{ji'},
\end{equation}
where $K^q_{ij}$ is the $(i,j)$ element of $\mathbf{K}(\phi_q)$. Large values of $[\Sigma_q]_{(i,j),(i',j')}$ occur when the two dyads have similar locations in either alignment (e.g., $i$ is near $i'$ and $j$ is near $j'$, or $i$ is near $j'$ and $j$ is near $i'$); small values occur when all four locations are spatially distant. 
The separable structure lets us represent $\boldsymbol{\Sigma}_q$ using an $n\times n$ node-level kernel rather than specifying its $\mathcal{O}(n^4)$ entries directly on $\mathcal{D}$, and it guarantees that the result is positive-definite. The resulting covariance is restricted to the index set $\{i<j\}$ using sparse matrix multiplication. 
We derive the updates for $\mathbf{w}_q$ and $\phi_q$ and quantify their cost in Appendices E and H, respectively.

The formulation in (\ref{eq:dyad-cov}) measures similarity between the first nodes of the two dyads ($K_{ii'}^q$), between the second nodes ($K_{jj'}^q$), and across the two positions ($K_{ij'}^qK_{ji'}^q$). 
The second term is what makes $\boldsymbol{\Sigma}_q$ invariant to which member of a pair is labeled first, which is required because the pairs are unordered.
Such a dyadic covariance structure is relevant for invasion dynamics because it allows for similarity among locations that produce or receive many migrants, such as points of introduction and areas of suitable habitat, as well as the linkage between two such locations. Invasions depend on these three concepts: where invasive individuals are produced/introduced, where they can establish, and how well those places are connected.

We construct $\mathbf{K}(\phi_q)$ with the Matérn 3/2 correlation function:
\begin{equation}\label{eq:K}
    [K(\phi_q)]_{i,j}=\left(1+\frac{\sqrt{3}d_{i,j}}{\phi_q}\right)\exp\left(-\frac{\sqrt{3}d_{i,j}}{\phi_q}\right)\quad \text{for }i,j\in\{1,\dots,n\},
\end{equation}
where $\phi_q$ is the spatial range parameter and $d_{i,j}=|\mathbf{s}_i-\mathbf{s}_j|$ is the Euclidean distance between $\mathbf{s}_i$ and $\mathbf{s}_j$.
We specify $\mathbf{K}(\phi_q)$ as a spatial correlation matrix and rescale (\ref{eq:dyad-cov}) to unit diagonal by dividing the entry for dyads $(i,j)$ and $(i',j')$ by $\{(1+(K^q_{ij})^2)(1+(K^q_{i'j'})^2)\}^{1/2}$. Doing so fixes the marginal scale of $\mathbf{W}$ (i.e., unit marginal variance), which regularizes the decomposition and improves sampling efficiency.
We model spatial dependency with the Matérn 3/2 kernel because it 
is smooth enough to accommodate gradual spatial variation and rough enough to represent localized structure \citep{hoeting2006model}.

The loading matrix $\mathbf{C}$ maps the latent spatial factors onto the dyadic coefficients. 
To stabilize the factor decomposition and encourage interpretability, we specify global-local shrinkage priors for $\mathbf{C}$ that concentrate posterior mass on a parsimonious set of loadings and suppress redundant factors. In particular, each column $\mathbf{c}_q$ of $\mathbf{C}$ has prior covariance $\mathbf{D}_q$ in an automatic relevance determination-type prior \citep{oba2003bayesian}:
$\mathbf{c}_q\sim\text{N}(\boldsymbol0,\mathbf{D}_q),$
where $\mathbf{D}_q=\text{diag}(\lambda^2_{lq}\xi^2_q)$. We used a half-Cauchy prior for the local scales $\{\lambda_{lq}\}$ and the global scales $\{\xi_q\}$:
$\lambda_{lq},\xi_q\sim \text{C}^+(0,1)$ for $l=1,\dots,p$ and $q=1,\dots,Q$.
Thus, each row of $\mathbf{C}$ has local shrinkage parameters $\{\lambda_{lq}\}_{q=1}^Q$ that minimize the contribution of redundant loadings, and each column of $\mathbf{C}$ has a global shrinkage parameter $\xi_q$ that suppresses entire factors if necessary. Such a hierarchical global-local shrinkage prior belongs to the horseshoe family, which adaptively enforces sparsity while preserving the possibility of large, interpretable effects \citep{carvalho2009handling}. 
We set $Q$ as the number of covariates in the model and let the shrinkage prior decide how much of that space to use; we report the resulting effective rank and the cost of setting $Q$ too small or too large in Appendix C.

Our specification for $\mathbf{C}$ is appropriate for landscape genomics because the number of candidate covariates are often large relative to the information available in the dyads; therefore, only a small subset of environmental features is expected to exhibit strong spatially varying effects. 
In conjunction with the spatial factor specification, the proposed shrinkage framework allows DSVCs to accommodate complex, nonstationary, and directional effects without overfitting. Thus, the model highlights the most influential environmental effects of invasive spread while maintaining computational tractability and ecological interpretability.

Similar to other factor models, $\mathbf{W}$ and $\mathbf{C}$ are not uniquely identifiable because different rotations, sign changes, and column rescalings yield the same decomposition \citep{lopes2004bayesian,ren2013hierarchical,papastamoulis2022identifiability}. 
In particular, for any invertible $\mathbf{A}\in\mathbb{R}^{Q\times Q}$, both $\mathbf{WA}^{-1}$ and $\mathbf{CA}'$ yield the same $\boldsymbol{\Delta}$. As a result,  individual latent factors and their corresponding spatial range parameters are not identifiable. Therefore, we do not report convergence diagnostics for $\phi_q$.
However, $\boldsymbol\Delta$ and all predictive quantities are uniquely determined \citep{zhang2022spatial}; therefore, our inference focuses on $\boldsymbol\Delta$ rather than the individual factor matrices.
We summarize the complexity of the fitted coefficient surface by the singular values of the posterior mean of $\boldsymbol{\Delta}$, which are invariant to the transformation above and provide an effective rank for the surface. We also report the singular value spectrum to state how many latent processes the data support. 

\subsection{Connectivity Covariates}
\label{sec:CC}

Invasive spread is often affected by natural and anthropogenic pathways (e.g., rivers, roads, railroads), which can alter the effective resistance between locations. For example, aquatic pathways, such as rivers, have been found to decrease resistance distance for some species \citep{erHos2019landscape} and increase resistance distance for others \citep{bartakova2015terrestrial}. In the former case, the pathway facilitates gene flow (i.e., a corridor), whereas in the latter case, the pathway impedes gene flow (i.e., a barrier). 
Unlike the environmental covariates in $\widetilde{\mathbf{x}}_{ij}$, which exist continuously across space, pathway features are discrete and spatially localized, requiring a different representation. 
We treat pathways as connectivity covariates that allow us to characterize their heterogeneous influence on invasion.

Conceptually, the proposed connectivity covariates aim to score how well two observations are connected by discrete landscape features. As an example, we consider two genotypes. If sampled beside the same river, they should receive a high score. However, two genotypes sampled beside different rivers (or far from any river) should receive a low score.
Connectivity covariates are used to determine whether genetic dissimilarity is influenced by shared access to a discrete pathway, as opposed to environmental covariates, which are used to determine whether genetic dissimilarity is affected by the environmental contrast between two locations.

\begin{figure}
    \centering
    \includegraphics[width=\linewidth]{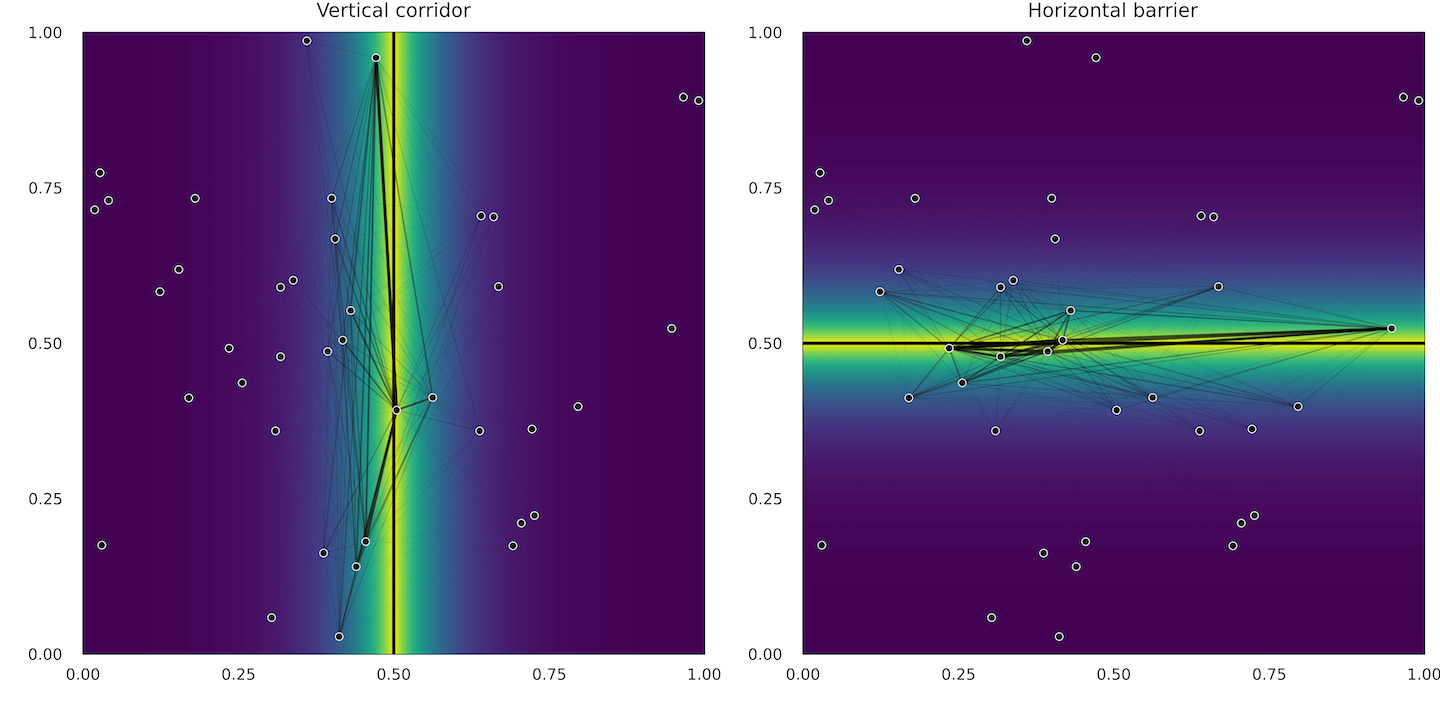}
    \caption{Closeness surfaces for the vertical corridor (left) and horizontal barrier (right) of the simulation study with pairwise shared‐segment scores superimposed as network edges whose width and opacity scale with shared‐segment scores. Black dots denote observations. Closeness and shared-segment scores do not encode whether the landscape features are barriers or corridors; this is handled by the corresponding regression coefficient.}
    \label{fig:sim-fig-1}
\end{figure}

Mathematically, we consider $C$ classes of pathways, each containing all landscape features of a certain type (e.g., rivers). For each class $c=1,\dots,C$, we compute closeness scores between each observed location and each feature in the class via an element-wise exponential decay transform. In particular, the closeness scores for node $i$ and class $c$ are
\begin{equation*}
    \mathbf{v}_{i,c}=\exp(-\mathbf{d}_{i,c}/\tau_c),
\end{equation*}
where $\mathbf{d}_{i,c}\in\mathbb{R}^{n_c}$ comprises the distances from node $i$ to each of the $n_c$ features in class $c$ and $\tau_c>0$ controls the rate of decay in the closeness measure for class $c$. We use exponential decay to quantify closeness because the influence of a pathway diminishes rapidly with distance; a particular river segment close to a site has strong influence, whereas distant river segments have negligible effect \citep{ward2002landscape}. Exponential decay provides a smooth, differentiable weighting that avoids arbitrary cutoffs and allows sites to be influenced by multiple nearby features at varying levels.
We let $\mathbf{V}_c\in\mathbb{R}^{n\times n_c}$ denote the closeness scores between all node-level observations and features in class $c$, where the $i$th row of $\mathbf{V}_c$ is $\mathbf{v}_{i,c}'$.
Using the closeness scores, we compute shared-segment scores $\mathbf{S}_c=\mathbf{V}_c\mathbf{V}_c',$ which quantify nearby connectivity features among sites; if two sites are close to the same pathway, their shared-segment score will be relatively large. We normalize the shared-segment scores  to ensure that they are on comparable scales across the $C$ classes: $\mathbf{S}^*_c=\frac{1}{n_c}\mathbf{S}_c$, where $n_c$ denotes the number of features in class $c$. Finally, we extract the upper-triangular entries of $\mathbf{S}^*_c$ such that the connectivity covariates align with the index set $\{i<j\}$. 
We let $\kappa_{ij,c}=[\mathbf{S}^*_c]_{ij}$ denote the connectivity covariate for dyad $(i,j)$ and class $c$. Each dyad gets $C$ connectivity covariates that represent shared proximity to pathways: $\boldsymbol\kappa_{ij}=(\kappa_{ij,1},\dots,\kappa_{ij,C})'.$
Because $\mathbf{S}_c$ is symmetric, the connectivity covariates contribute to the even component of the mean structure and cannot be used to make asymmetric inference.
We provide an example of closeness scores across space and shared-segment scores between observations in Figure \ref{fig:sim-fig-1}. Pairs of nodes have higher shared-segment scores (i.e., thicker, opaque edges) when they are both near common landscape features.

The decay parameters $\{\tau_c\}$ determine the spatial scale at which a pathway class is shared between two nodes. As $\tau_c\to0$, the connectivity covariate approaches zero for every dyad, and as $\tau_c\to\infty$, it approaches a constant that the intercept absorbs. We specify the prior $$\log\tau_c\sim\text{N}(\log\tau^0_c,\sigma^2_{\log\tau}),$$
where $\tau_c^0=\text{median}(\mathbf{d}_{c})/\log(2)$, $\mathbf{d}_{c}$ comprises the distances from each node to the nearest feature in class $c$, and $\sigma^2_{\log\tau}$ is the prior variance.
We set $\sigma_{\log\tau}=\log(4)/2$ such that the central $95\%$ of the prior spans $[0.25\tau_c^0,4\tau_c^0]$.
Finally, we measure the number of relevant features with the participation ratio $(\sum_jv_{ij,c})^2/\sum_jv_{ij,c}^2$, which counts the features that effectively contribute to the closeness score of node $i$. We report the participation ratio for each class and discuss our choice for nearest-feature distance in Appendix C.


The analysis of discrete pathways has a rich history in landscape genomics. For example, circuit-theoretic methods treat rasters containing pathways as electrical resistors and compute effective resistance between nodes \citep{mcrae2006isolation}. Such rasters can incorporate barriers (high resistance) or corridors (low resistance). The proposed connectivity covariates can be seen as the dyadic analogue to pathway rasters in circuit-theoretic methods. However, instead of encoding resistance in a continuous raster, we represent discrete pathways (e.g., roads, rivers) as covariates in a dyadic regression. As a result, we can explicitly test the role of individual connectivity classes rather than inferring them indirectly from raster resistances. Additionally, because the connectivity covariates are defined in dyadic space, they integrate seamlessly into the DSVC framework and can be analyzed alongside environmental and spatial predictors. Incorporating the connectivity covariates extends our regression model:
\begin{equation}\label{eq:kappa-SVCs}
    \widetilde{y}_{ij}\sim \text{N}\left(\alpha+\mathbf{z}_{ij}'(\boldsymbol{\beta} + \boldsymbol\delta_{ij}) + \widetilde{\eta}_{ij},\sigma^2\right),
\end{equation}
where $\mathbf{z}_{ij}=[\mathbf{\widetilde{x}}'_{ij} \; \boldsymbol\kappa'_{ij}]'$ is the augmented dyadic design matrix with pairwise differenced environmental covariates $\widetilde{\mathbf{x}}_{ij}$ and connectivity covariates $\boldsymbol\kappa_{ij}$, $\boldsymbol\beta\in\mathbb{R}^{p+C}$ are the global dyadic coefficients, and $\boldsymbol\delta_{ij}\in\mathbb{R}^{p+C}$ are the DSVCs at the pairwise location set $(\mathbf{s}_i,\mathbf{s}_j)$. 
A negative coefficient for a connectivity covariate suggests that the pathway class acts as a corridor, whereas a positive coefficient suggests that it acts as a barrier. 

We model connectivity covariates with DSVCs to allow the effects of pathways to vary across space. For example, rivers might serve as corridors in upstream sections but as barriers downstream \citep{bartakova2015terrestrial,erHos2019landscape}, and highways may facilitate spread near urban centers but play little role in remote areas \citep{medley2015human}. Coupling DSVCs with connectivity covariates provides a principled way to account for spatial heterogeneity and to identify localized effects of pathways.
Alternatively, connectivity covariates can be modeled using only fixed effects, which may be more appropriate when the pathway data lack sufficient spatial resolution to justify local variation. 

\section{Implementation}
\label{sec:implementation}

We fit the proposed Bayesian hierarchical model (described in full in Appendix D) using a blocked Gibbs-Metropolis sampler. The sampler is structured as four blocks: (i) the node-level latent field $\boldsymbol{\eta}$ and its spatial range, (ii) the latent dyadic spatial factors $\mathbf{W}$ and their ranges, (iii) the loading matrix $\mathbf{C}$, and (iv) the decay parameters $\{\tau_c\}$. We describe each in order; full derivations can be found in Appendix E.

To avoid confounding with the intercept, we updated $\boldsymbol{\eta}$ in the $(n-1)$-dimensional subspace orthogonal to $\boldsymbol{1}_n$, where the conditional update is conjugate and the covariance remains positive-definite at long spatial ranges.
Because the latent field has $n-1$ effective degrees of freedom, $\sigma^2_\eta$ and $\phi_\eta$ are poorly identified and are weakly separated from each other. We conduct a prior sensitivity analysis for $\sigma^2_\eta$ in Appendix C.
For the spatial range parameter, we updated $\phi_\eta$ via slice sampling on $\log\phi_\eta$ with ``stepping out,'' which avoids the manual tuning of proposal variances given the initial width parameter, and regularized the update with an informative prior centered on the median inter-node distance \citep[][]{neal2003slice}.

We updated $\phi_q$ similar to $\phi_\eta$ and alongside $\mathbf{W}$. To stabilize MCMC sampling, we alternated between (i) a whitened joint random-walk proposal for $(\phi_q,\mathbf{w}_q)$ and (ii) the conditional update for $\mathbf{w}_q\mid\phi_q$, which balances mixing of the spatial range against exact updates of the factors \citep[][]{van2015metropolis}. 

We drew each column of the loading matrix from a conjugate Gaussian update with a Cholesky precision solve under the Makalic-Schmidt inverse-gamma augmentation of the half-Cauchy shrinkage scales \citep{makalic2015simple}. After updating $\mathbf{C}$, we recentered $\mathbf{W}$, shifted constants into $\boldsymbol{\beta}$, and reset $\boldsymbol{\Delta}=\mathbf{W}\mathbf{C}'$; this procedure regularized the decomposition $\boldsymbol\Delta=\mathbf{WC}'$ by fixing the marginal scale of $\mathbf{W}$ and pushing global signal into $\boldsymbol\beta$.

Every tenth iteration, we updated each decay parameter $\tau_c$ with a random-walk Metropolis step on $\log\tau_c$ using a proposal standard deviation that was adapted during burn-in toward an acceptance rate of $0.30$. Each proposal recomputed the closeness and shared-segment scores for one class and one column of the dyadic design matrix; we provide the cost of this update in Appendix E. Then, we centered and scaled the connectivity covariates at the current value of $\tau_c$ such that $\beta_{p+c}$ was the effect of a one standard deviation increase in the shared-segment score of class $c$ at that decay.

For computational efficiency, we implemented the MCMC algorithm in Julia \citep{bezanson2018julia}.
We ran our MCMC algorithm for 15,000 and 50,000 iterations in the simulation study and cheatgrass analysis, respectively, with a burn-in period of 20\% of the iterations and thinning every fifth iteration in the simulation study and every tenth iteration in the cheatgrass analysis. The computational costs of our model are provided in the respective sections.
Convergence was assessed by split potential scale reduction factor $\widehat{R}$ and effective sample size \citep[ESS;][]{gelman1995bayesian} for $\alpha$, $\boldsymbol{\beta}$, $\boldsymbol{\Delta}$, $\boldsymbol{\eta}$, $\sigma^2$, $\sigma^2_\eta$, $\phi_\eta$, and $\{\tau_c\}$.

\section{Simulation Study}

We simulated dyadic data under three data-generating processes (s1, s2, and s3) and aimed to recover the true parameter values in each; the three processes differed in whether the model was correctly specified. Across 25 independent replicates per process, we sampled $n=40$ nodes uniformly on $\mathcal{S}=[0,1]\times[0,1]$ with $p=4$ node-level covariates and specified two pathways: a horizontal barrier and a vertical corridor. Each pathway made up its own class (i.e., $C=2$). We provide the full design details in Appendix F. 

Under (s1), the model was correctly specified with $\boldsymbol{\Delta}$ drawn from the latent factor prior at $Q=6$. Under (s2), $\boldsymbol{\Delta}$ was misspecified and was constructed from a deterministic step field over dyad midpoints, which had no low-rank representation, was not Gaussian, and was discontinuous in space. Under (s3), we removed the DSVCs ($\boldsymbol{\Delta}=\boldsymbol{0}$) and let the dyadic errors follow a $t_5$ distribution; this model explored whether the added flexibility in the proposed framework was problematic when the flexibility was unnecessary.

We fixed $\alpha=-0.81$, $\sigma^2=0.150$, $\sigma^2_\eta=0.019$, and $\phi_\eta=\max(\mathbf{D}_{\text{euc}})/5$ at the posterior means from the full invaded-range fit in Section \ref{sec:cheatgrass}, and we set $\boldsymbol{\beta}=(-0.478, 0.081, -0.226, \\ -0.084, -0.274, 0.015)'$. 
Then, we simulated $\boldsymbol{\eta}\sim\text{N}(\boldsymbol{0},\sigma^2_\eta\mathbf{R}(\phi_\eta))$, where $\mathbf{R}(\phi_\eta)$ was an exponential correlation kernel with spatial range $\phi_\eta$. 
We centered $\boldsymbol{\Delta}$ to ensure that the DSVCs were not confounded with $\boldsymbol{\beta}$ and rescaled $\boldsymbol{\Delta}$ under (s1) and (s2) to a common root mean square of $0.30$, which put the correctly specified and the misspecified process at equal signal strength.

We fit four versions of our Bayesian hierarchical model to each process: (i) the standard dyadic model in \eqref{eq:mini} without connectivity covariates and DSVCs, (ii) the model in \eqref{eq:kappa-SVCs} without DSVCs but with connectivity covariates, (iii) the model in \eqref{eq:kappa-SVCs} without connectivity covariates but with DSVCs, and (iv) the full model in \eqref{eq:kappa-SVCs} with both DSVCs and connectivity covariates. We fit each model using 15,000 iterations with a burn-in of 3,000 iterations and thinning every five iterations. The median sampling rates of models (i) and (iv) under (s1) were 3186 it/s and 17.6 it/s, respectively, on a 3.2 GHz processor with 64 GB of RAM.

For $\boldsymbol{\beta}$, we report root mean square error (RMSE) and $95\%$ interval coverage in Table \ref{tab:sim-main}, and we report the bias and coverage of $\alpha$, $\sigma^2$, $\sigma^2_\eta$, and $\phi_\eta$ in Appendix F. For the coefficient surface $\boldsymbol{\Delta}$, we report integrated RMSE and coverage. For predictive comparison, we report continuous ranked probability scores \citep[CRPS; ][]{matheson1976scoring} and widely applicable information criterion \citep[WAIC; ][]{watanabe2010asymptotic}. We also provide Monte Carlo standard errors across the 25 replicates in Table \ref{tab:sim-main}.

\begin{table}
\caption{Replicated simulation study. Each cell is the mean over replicates with the Monte Carlo standard error in parentheses. CRPS is computed on the logit scale. We do not mark the best performing model in terms of $\boldsymbol{\Delta}$ in (s3) because the true surface is $\boldsymbol{0}$, and a model containing no coefficient surface attains a $\boldsymbol{\Delta}$ RMSE of zero by construction.}
\label{tab:sim-main}
\centering\footnotesize
\setlength{\tabcolsep}{4pt}
\resizebox{\textwidth}{!}{%
\begin{tabular}{llrrrrrr}
\toprule
Data Generating Process & Model & $\boldsymbol{\beta}$ RMSE & $\boldsymbol{\beta}$ coverage & $\boldsymbol{\Delta}$ RMSE & $\boldsymbol{\Delta}$ coverage & CRPS  & WAIC \\
\midrule
\multirow{4}{*}{(s1): correct} & (i) standard & 0.059 (0.005) & 0.95 & 0.305 (0.019) & --- & 0.540 (0.014) & 2299.3 \\
 & (ii) + connectivity & 0.099 (0.013) & 0.81 & 0.300 (0.000) & --- & 0.509 (0.016) & 2174.0 \\
 & (iii) + DSVCs & \textbf{0.031 (0.003)} & 0.95 & 0.168 (0.012) & 0.96 & 0.230 (0.007) & 1438.1 \\
 & (iv) full & 0.040 (0.004) & 0.95 & \textbf{0.141 (0.010)} & 0.96 & \textbf{0.188 (0.004)} & \textbf{1069.4} \\
\midrule
\multirow{4}{*}{(s2): misspecified $\boldsymbol{\Delta}$} & (i) standard & 0.066 (0.003) & 0.91 & 0.300 (0.000) & --- & 0.486 (0.006) & 2110.6 \\
 & (ii) + connectivity & 0.096 (0.007) & 0.74 & 0.300 (0.000) & --- & 0.468 (0.006) & 2038.4 \\
 & (iii) + DSVCs & 0.029 (0.002) & 0.95 & 0.137 (0.004) & 0.94 & 0.240 (0.007) & 1455.2 \\
 & (iv) full & \textbf{0.028 (0.002)} & 0.96 & \textbf{0.110 (0.002)} & 0.95 & \textbf{0.185 (0.003)} & \textbf{996.9} \\
\midrule
\multirow{4}{*}{(s3): no DSVCs, $t_5$ errors} & (i) standard & 0.024 (0.002) & 0.96 & 0.000 (0.000) & 1.00 & 0.242 (0.001) & 1056.3 \\
 & (ii) + connectivity & \textbf{0.016 (0.001)} & 0.98 & 0.000 (0.000) & 1.00 & 0.208 (0.001) & 770.3 \\
 & (iii) + DSVCs & 0.023 (0.002) & 0.95 & 0.030 (0.004) & 0.99 & 0.215 (0.004) & 1019.3 \\
 & (iv) full & 0.017 (0.001) & 0.97 & 0.010 (0.001) & 1.00 & \textbf{0.197 (0.002)} & \textbf{769.0} \\
\midrule
\bottomrule\end{tabular}}\end{table}

Under (s1) and (s2), the models without DSVCs failed to recover the coefficient surface because they contained no such term, and their estimate of the dyadic variance absorbed the omitted surface. 
The two models containing DSVCs recovered the surface at coverage 0.96 under (s1) and 0.94 under (s2), and the full model reduced the bias in the dyadic variance to 0.014 and 0.001, respectively. Coverage of the global coefficients was at its nominal level (95\%) for every model except the connectivity-only model, which only reached 0.81 under (s1) and 0.74 under (s2).
Omitting a nonstationary surface while retaining connectivity covariates did not only cost precision; the connectivity coefficients absorbed the omitted structure, and their intervals consequently failed to cover the values that generated the data.

Under (s3), the models containing DSVCs did not degrade. Coverage of the global coefficients was 0.95 for the DSVC-only model and 0.97 for the full model, and the fitted surface was near zero with integrated RMSEs of 0.03 and 0.01, respectively. Therefore, the DSVCs cost little when the process that motivates them was absent. 

In Appendix C, we report two sensitivity analyses. First, we set the factor dimension below the true rank, which cost coverage of the coefficient surface. We found that setting $Q$ above the true rank cost sampling time without improving recovery. Thus, the global-local shrinkage prior protected against a factor space that was too large and did not fix one that was too small. 
Second, we fit the models containing connectivity covariates under (s1) with the true decay multiplied by 0.5, 1, and 2, which measured how much the fit degraded when the decay was poorly fixed.
We found that the fit was insensitive to the decay over $[0.5\tau_c^*,2\tau_c^*]$, where $\tau_c^*$ was the true value.

\section{Cheatgrass Case Study}
\label{sec:cheatgrass}

Our cheatgrass dataset comprised spatially-referenced genotype sequences throughout portions of the invaded range (Intermountain West) and native range (Central Europe). For each range, we generated the corresponding genetic distance matrix (GDM) $\mathbf{D}_{r}\in\mathbb{N}_0^{n_r\times n_r}$ for $r\in\{\text{native, invaded}\}$, where $n_\text{invaded}=86$ and $n_\text{native}=105$; we outline this process in Appendix A. The $(i,j)$ element of the GDM corresponded to the number of allele-sharing mismatches (i.e., discordant loci) between the $i$th and $j$th individuals. In the invaded range, 28 of the 3655 dyads had a kinship above 0.995; we refer to these dyads as near-clonal pairs throughout our analysis. We computed the dyadic response $\widetilde{y}_{ij}$ between individuals $i$ and $j$ as the logit of the per-dyad mismatch proportion in (\ref{eq:pij}), where $M_{ij}=2100$ for every pair.

For the node-level environmental covariates, we extracted and standardized the nineteen CHELSA bioclimatic variables at the genotype locations \citep{karger2017climatologies}, appended longitude and latitude, computed the pairwise differences, and applied a radial basis function (RBF) transformation to represent potential nonlinear relationships between covariates and gene flow. We selected five centers by $k$-means clustering on $\widetilde{\mathbf{x}}_{ij}$ and set the kernel bandwidth from the median inter-center distance; we report sensitivity to the number of centers in Appendix C. The transformation returned $\widetilde{\mathbf{x}}_{ij}\in\mathbb{R}^5$, where $p=5$.


In the invaded range, we considered $C=3$ pathway classes: historic railroads, modern highways, and rivers/streams. Railroads have long been hypothesized to have facilitated the initial spread of cheatgrass across North America by enabling westward expansion \citep{mack1981invasion}. As a result, livestock historically concentrated around rail lines in the West, and livestock are known to facilitate cheatgrass dispersal \citep{williamson2020fire}. By including historic rail lines as a pathway class, we tested whether historic corridors of movement continue to leave a detectable imprint on modern population structure, even though their historic role in expansion and livestock concentration has diminished.
We also investigate modern highways as a pathway class to contrast with historic railroads. Historic rail lines reflect the sparse corridors that structured the initial spread of cheatgrass, whereas highways form a dense network that could affect ongoing dispersal. Additionally, highways have been comparatively understudied as potential pathways for cheatgrass dispersal. 
Finally, we included rivers and streams because wildlife and livestock grazing is often concentrated along riparian corridors, and movement of animals along these pathways is hypothesized to be an important mechanism for cheatgrass dispersal \citep{king2019potential,molvar2024cheatgrass}. We obtained 2, 20, and 745 features in the historic railroads, modern highways, and rivers/streams classes, respectively, from the North American Environmental Atlas \citep{cec-data}. For each class, we computed the closeness scores $\mathbf{V}_c$ between each observed genotype and each feature in the class, then computed the pairwise shared-segment scores $\mathbf{S}_c=\mathbf{V}_c\mathbf{V}_c'$. Finally, we normalized the shared-segment scores and extracted the upper-triangular entries to obtain the connectivity covariates  $\boldsymbol\kappa_{ij}\in\mathbb{R}^3$ for the three pathway classes. See Appendix G for a plot of $\mathbf{V}_c$ across the invaded range for each class.

We fit four versions of our Bayesian hierarchical model to the cheatgrass data in the invaded range and two in the native range (see Table \ref{tab:t1}). In the native range, we fit (n1) the standard dyadic model and (n2) the dyadic model with DSVCs. Comparing (n1) and (n2) tests whether nonstationary environmental responses are necessary to explain gene flow under long-term equilibrium. 
In the invaded range, we fit (i1) the standard dyadic model, (i2) the model with connectivity covariates alone, (i3) the model with DSVCs alone, and (i4) the full dyadic model with DSVCs and connectivity covariates. 
By fitting (i2) and (i3), we are able to attribute the improvement to each component separately.
By comparing all four model fits, we can assess whether invasion dynamics are driven by heterogeneous environmental effects or anthropogenic and ecological pathways of spread. Thus, the comparison reveals whether different processes govern gene flow in native versus invasive populations.

\begin{table}
\caption{Model comparison in the native and invaded range. CRPS is computed on the logit scale. WAIC is computed from the Gaussian log density of the logit-scale response; it is negative when the dyadic variance is small. Bold marks the best performing model within each range.}
\label{tab:t1}
\begin{tabular}{llrrrr}
\toprule
Range & Model & CRPS & $R^2$ & WAIC & Runtime (it/s) \\
\midrule
\multirow{2}{*}{native} & (n1) standard & 0.1247 & 0.427 & 1142.5 & 416 \\
 & (n2) + DSVCs & \textbf{0.0235} & \textbf{0.982} & $\mathbf{-12842.3}$ & 0.59 \\
\midrule
\multirow{4}{*}{invaded} & (i1) standard & 0.2689 & 0.115 & 6749.7 & 690 \\
 & (i2) + connectivity & 0.2629 & 0.175 & 6674.5 & 638 \\
 & (i3) + DSVCs & 0.1771 & 0.710 & 5198.3 & 1.42 \\
 & (i4) full & \textbf{0.1555} & \textbf{0.816} & \textbf{4914.7} & 1.08 \\
\bottomrule
\end{tabular}
\end{table}

\subsection{Analysis in the Native Range}

For (n1), the posterior mean of the intercept was $\hat\alpha=-0.7545$ (95\% CI $[-0.7774,-0.7310]$, ESS$=1947$). Chains for the global regression coefficients $\boldsymbol\beta$ converged with ESS ranging from 2516 to 3672, and four of the five had credible intervals excluding zero. Chains for the latent spatial random effects $\boldsymbol\eta$ converged with ESS ranging from 3400 to 4000. The posterior mean of the variance of the spatial random effects was $\hat\sigma^2_\eta=0.0200$ (95\% CI $[0.0073,0.0533]$, ESS$=127$), and the posterior mean of the dyadic variance was $\hat\sigma^2=0.0709$ (95\% CI $[0.0683,0.0736]$, ESS$=3845$). The lowest ESS and the highest $\hat{R}$ occurred for the spatial range parameter $\phi_\eta$ (ESS$=110$, $\widehat{R}=1.006$).

For (n2), 
the posterior mean of $\sigma^2_\eta$ was $\hat\sigma^2_\eta=0.0302$ (95\% CI $[0.0108,0.0755]$, ESS$=146$), the posterior mean of $\sigma^2$ was $\hat\sigma^2=0.0036$ (95\% CI $[0.0033,0.0038]$, ESS$=602$), and the chains for $\boldsymbol{\eta}$ and for the DSVCs converged with ESS ranging from 102 to 393 and from 35 to 135, respectively.
A scale invariance of the latent factorization admitted a ridge along which one latent factor collapsed, and the fifth factor of (n2) traversed that ridge; we discuss this further in Appendix C.

The CRPS for (n1) and (n2) were 0.1247 and 0.0235, respectively, and the coefficient of determination across observed pairs was 0.427 and 0.982, respectively.
Cheatgrass populations in the native range have had long-term exposure to local environmental gradients, and stable genotype-environment associations should leave most of the signal in the global environmental coefficients. However, accounting for nonstationary environmental effects further improved the model fit. The DSVCs absorbed 94.9\% of the dyadic variance in the native range, leaving the residual dyadic variance at 0.0036.

\subsection{Analysis in the Invaded Range}

Models (i2) and (i3) isolated the contribution of the connectivity covariates and the DSVCs, respectively. The connectivity covariates alone improved WAIC over (i1) by 75, and DSVCs alone improved WAIC by 1551. Together, connectivity covariates and DSVCs improved WAIC by 1835, which exceeded the sum of the separate improvements; this pattern was consistent for CRPS and $R^2$. Thus, DSVCs and connectivity covariates were complementary rather than redundant. 
Together, the proposed framework allowed us to infer how pathways facilitated or impeded gene flow to a degree that itself varied across the range.

For (i1), the posterior mean of the intercept was $\hat\alpha=-0.6894$ (95\% CI $[-0.7364,-0.6397]$, ESS$=3029$). Chains for the global regression coefficients $\boldsymbol{\beta}$ converged with ESS ranging from 3710 to 4000, and three of the five had credible intervals excluding zero. 
Chains for the latent spatial random effects $\boldsymbol\eta$ converged with ESS ranging from 3409 to 4000.
The posterior mean of $\sigma^2_\eta$ was $\hat\sigma^2_\eta=0.0294$ ($95\%$ CI $[0.0127,0.0630]$, ESS $=235$). The posterior mean of $\sigma^2$ was $\hat\sigma^2=0.3642$ ($95\%$ CI $[0.3478,0.3811]$, ESS $=3854$). The lowest ESS occurred for the spatial range parameter $\phi_\eta$ (ESS $=168$), which also had the largest $\hat{R}$ value ($\hat{R}=1.009$).  

For (i2), the posterior mean of the intercept was $\hat\alpha=-0.7605$ ($95\%$ CI $[-0.8108,-0.7103]$, ESS$=2726$), the dyadic variance was $\hat{\sigma}^2=0.3568$ (95\% CI $[0.3407,0.3736]$), and the latent spatial variance was $\hat{\sigma}^2_\eta=0.0285$ (95\% CI $[0.0127,0.0618]$).
For (i3), the coefficient surface converged with median ESS above 2000.

For (i4), the posterior mean of the intercept was $\hat{\alpha}=-0.8069$ (95\% $[-0.8813,-0.7258]$, ESS$=437$).
Chains for the $\boldsymbol\beta$ coefficients converged with ESS ranging from 459 to 2147, only one of which had a 95\% credible interval excluding zero; this suggests that explanatory power was partially absorbed by the DSVC terms. The $\boldsymbol{\delta}$ chains converged with median ESS 2489 and maximum $\widehat{R}$ 1.005. 
The singular values of the posterior mean of $\boldsymbol{\Delta}$ gave an effective rank of two components carrying 94.4\% of the fitted surface and three carrying 99.0\%, which indicates that the global-local shrinkage prior collapsed the $Q=8$ factor space to a smaller number of active processes.
Chains for the latent spatial random effects $\boldsymbol{\eta}$ converged with ESS ranging from 1035 to 3550. 
The node-level spatial range parameter $\phi_\eta$ converged well (ESS $=338$, $\hat{R}=1.007$). The posterior mean of $\sigma^2$ was $\hat\sigma^2=0.1502$ ($95\%$ CI $[0.1401,0.1606]$, ESS $=2101$), revealing less variability than in (i1). Finally, the posterior mean for the latent spatial variance was $\hat\sigma^2_\eta=0.0191$ ($95\%$ CI $[0.0090,0.0385]$, ESS $=966$). Model (i4) produced lower estimates of both $\sigma^2$ and $\sigma^2_\eta$ than any other invaded-range fit, indicating that the DSVCs and the connectivity covariates together reduced unexplained variability in the invaded range. 

Relative to the standard dyadic model (i1), the full dyadic model (i4) reallocated explanatory power from the global coefficients $\boldsymbol\beta$ to spatially varying effects as evidenced by the reduced number of significant $\boldsymbol{\beta}$ estimates and the role of $\boldsymbol{\delta}$ across dyadic space. 
We visualize the effect of the DSVCs across the Intermountain West in Appendix G, which maps the average absolute $z$-scores of posterior predictive DSVCs, aggregated per grid node across neighboring dyads and covariates. In particular, we constructed a fine grid $\mathcal{G}$ across the domain with equal geographic distance between grid nodes, displayed with a Mercator projection. For each grid cell $g\in\mathcal{G}$, we summarized DSVC signal by 
$$\bar{z}_g=\frac{1}{\deg(g)(p+C)}\sum_{g'\in\mathcal{N}(g)}\sum^{p+C}_{l=1}\bigg|\frac{\widehat{\Delta}_{(g,g'),l}}{\widehat{\text{sd}}(\Delta_{(g,g'),l})}\bigg|,$$ 
where $\deg(\cdot)$ is the number of valid neighbors (up to 4), $\mathcal{N}(g)$ is the set of cardinal neighbors for cell $g$, $\widehat{\Delta}_{(g,g'),l}$ is the posterior predictive DSVC for dyad $(g,g')$ and the $l$th covariate, and $\widehat{\text{sd}}(\Delta_{(g,g'),l})$ is its posterior standard deviation. Regions of large $\bar{z}_g$ are those in which environmental effects on gene flow are spatially heterogeneous.

We summarized the posterior effects of the connectivity covariates locally and globally.
For each grid node $g\in\mathcal{G}$, we computed the local node-level slope
$$\theta_{g,c}=\gamma_c+\frac{1}{\deg(g)}\sum_{g'\in\mathcal{N}(g)}\Delta_{(g,g'),c},$$
where $\gamma_c=\beta_{p+c}$ is the global slope for pathway class $c$ and $\Delta_{(g,g'),c}$ is the DSVC of dyad $(g,g')$ for class $c$.
Globally, the centering step in the sampler leaves each column of $\boldsymbol{\Delta}$ with mean zero across the observed dyads, so the domain-wide effect of pathway class $c$ is $\gamma_c$. We report $\{\gamma_c\}$ for (i2) and (i4) in Table \ref{tab:conduit} with the decay parameters both fixed and estimated.

\begin{table}
\caption{Posterior means of connectivity coefficients in the invaded range with 95\% credible intervals. Negative values facilitate gene flow. An asterisk marks an interval excluding zero.}
\label{tab:conduit}
\centering\footnotesize
\setlength{\tabcolsep}{4pt}
\resizebox{\textwidth}{!}{%
\begin{tabular}{lcccc}
\toprule
 & \multicolumn{2}{c}{(i2) connectivity only} & \multicolumn{2}{c}{(i4) full model} \\
\cmidrule(lr){2-3}\cmidrule(lr){4-5}
Pathway class & fixed $\tau_c$ & estimated $\tau_c$ & fixed $\tau_c$ & estimated $\tau_c$ \\
\midrule
Railroad       & $-0.019$ ($-0.049$, $0.010$)  & $0.093$ ($0.068$, $0.120$)$^{*}$   & $0.015$ ($-0.062$, $0.095$) & $0.023$ ($-0.045$, $0.093$) \\
Highway        & $0.047$ ($0.019$, $0.078$)$^{*}$ & $-0.083$ ($-0.114$, $-0.049$)$^{*}$ & $0.038$ ($-0.048$, $0.122$) & $0.010$ ($-0.039$, $0.059$) \\
River / stream & $-0.111$ ($-0.135$, $-0.087$)$^{*}$ & $-0.053$ ($-0.077$, $-0.030$)$^{*}$ & $-0.064$ ($-0.129$, $0.006$) & $-0.072$ ($-0.178$, $0.029$) \\
\bottomrule
\end{tabular}}
\end{table}

Every interval in (i2) with estimated $\{\tau_c\}$ excluded zero, and each was narrower than the corresponding interval in (i4). When fixing the decay parameters, the sign of the connectivity coefficients reversed for the railroad and highway classes. Thus, in the absence of a nonstationary surface, the connectivity covariates absorbed spatial variation in gene flow that they did not generate, and this absorption overstated the attribution to a pathway class. Such sensitivity was only relevant to attribution rather than the fit of the model. Across three fixed decays, the dyadic variance in (i2) shifted by roughly 0.3\%. Therefore, the decay parameters did not change how well the model described the data; they changed which pathway class received credit. For example, the railroad class was the one in which the attribution was least stable because there were only two features and only 11 of the 86 sampled genotypes were within 25 km of either. Therefore, no decay value yielded a covariate that adequately measured proximity to a railroad. In contrast, the river class did not have this problem because there were 745 features.
We present the decay diagnostics in Appendix C and the maps of $\theta_{g,c}$ in Appendix G.

The CRPS for (i1) and (i4) were 0.2689 and 0.1555, respectively, indicating that (i4) provided a significantly better fit to the invaded range data.
To assess predictive performance, we examined the residuals between true kinship $k_{ij}=1-\text{logistic}(\widetilde{y}_{ij})$ and predicted kinship $\hat{k}_{ij}=1-\text{logistic}(\widetilde{y}^*_{ij})$, where $\widetilde{y}^*_{ij}$ was obtained from posterior predictive draws under each model. The residuals under (i4) were roughly half the magnitude of those under (i1), further indicating that the proposed framework improved on the standard dyadic flow model under the presence of spatially heterogeneous effects (Appendix G). 

Model (i1) failed to predict near-clonal kinship because near-clonal pairs violated the isolation-by-distance assumption that genetic dissimilarity increases with geographic distance, whereas the proposed model predicted them more accurately by allowing for localized, nonstationary effects. 
The RMSE on the near-clonal pairs fell from 0.257 under (i1) to 0.060 under (i4), and the bias fell from $-0.255$ to $-0.050$. However, 90\% prediction interval coverage on those pairs rose only from 0.00 to 0.11, whereas coverage across all dyads remained near 0.974. Therefore, the full model located the near-clonal pairs much better than the standard model yet remained overconfident about them because the intervals narrowed faster than the errors. A model that coherently recovers near-clonal pairs would require a component for the mechanism that produces them, such as recent common ancestry through multiple introductions of the same lineage, which the proposed framework can only accommodate partially through $\boldsymbol{\Delta}$.

Overall, (i4) had the lowest CRPS, smaller residuals across all pairs, and an improved fit for near-clonal pairs that violated IBD assumptions. Thus, the DSVCs and connectivity covariates improved our understanding of cheatgrass invasion dynamics and predictive accuracy of genetic dissimilarity. However, the larger improvement in the native range suggests that nonstationarity in the native range was spatially coherent, whereas a substantial part of the variation in the invaded range was not spatially structured.

\subsection{Mapping Gene Flow}

While estimates of regression coefficients, connectivity covariates, and latent spatial factors provide mechanistic insights of nonstationary dynamics, conservation biologists and ecologists ultimately seek to know where gene flow is most likely to occur, which directly informs surveillance, containment, and eradication.
Thus, we translated posterior parameter estimates into vector field representations that display corridors, barriers, and heterogeneous dispersal in geographic space.  

To construct vector field maps, we used posterior means of the model parameters to predict the expected genetic dissimilarity between each interior grid node and each of its four cardinal neighbors and took central finite differences of those predictions in longitude and latitude. The resulting vectors $\{(u_g,v_g):g\in\mathcal{G}\}$ defined a mean vector field that summarized the local magnitude and direction of gene flow across the landscape; Appendix G gives the predictive mean and the difference quotients.  
The vector field shows how environmental heterogeneity, connectivity pathways, and latent spatial processes jointly influenced invasion dynamics. 

\begin{figure}[ht]
  \centering
  \begin{subfigure}{0.495\textwidth}
    \centering
    \includegraphics[width=\textwidth]{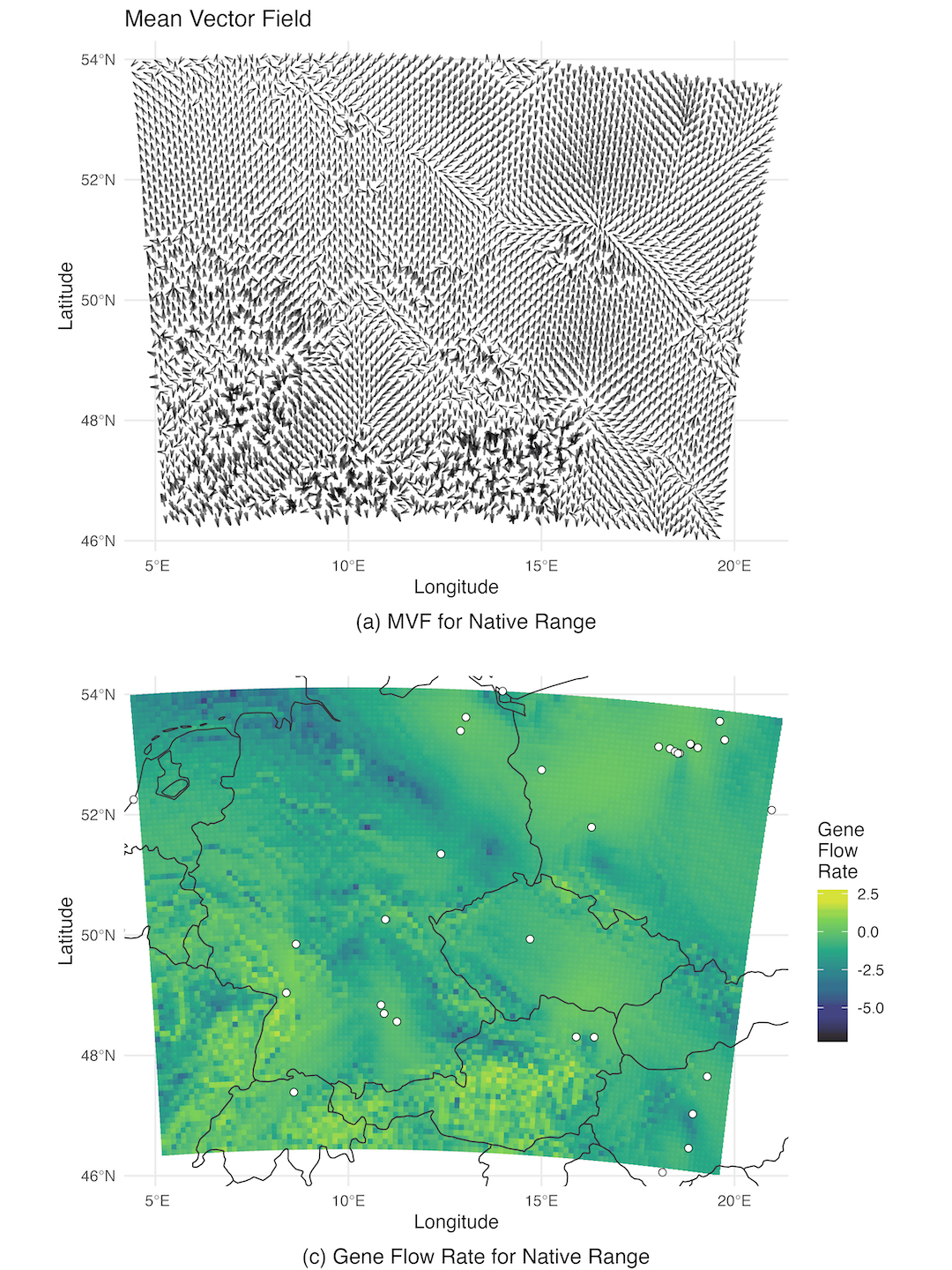}
  \end{subfigure}%
  \begin{subfigure}{0.495\textwidth}
    \centering
    \includegraphics[width=\textwidth]{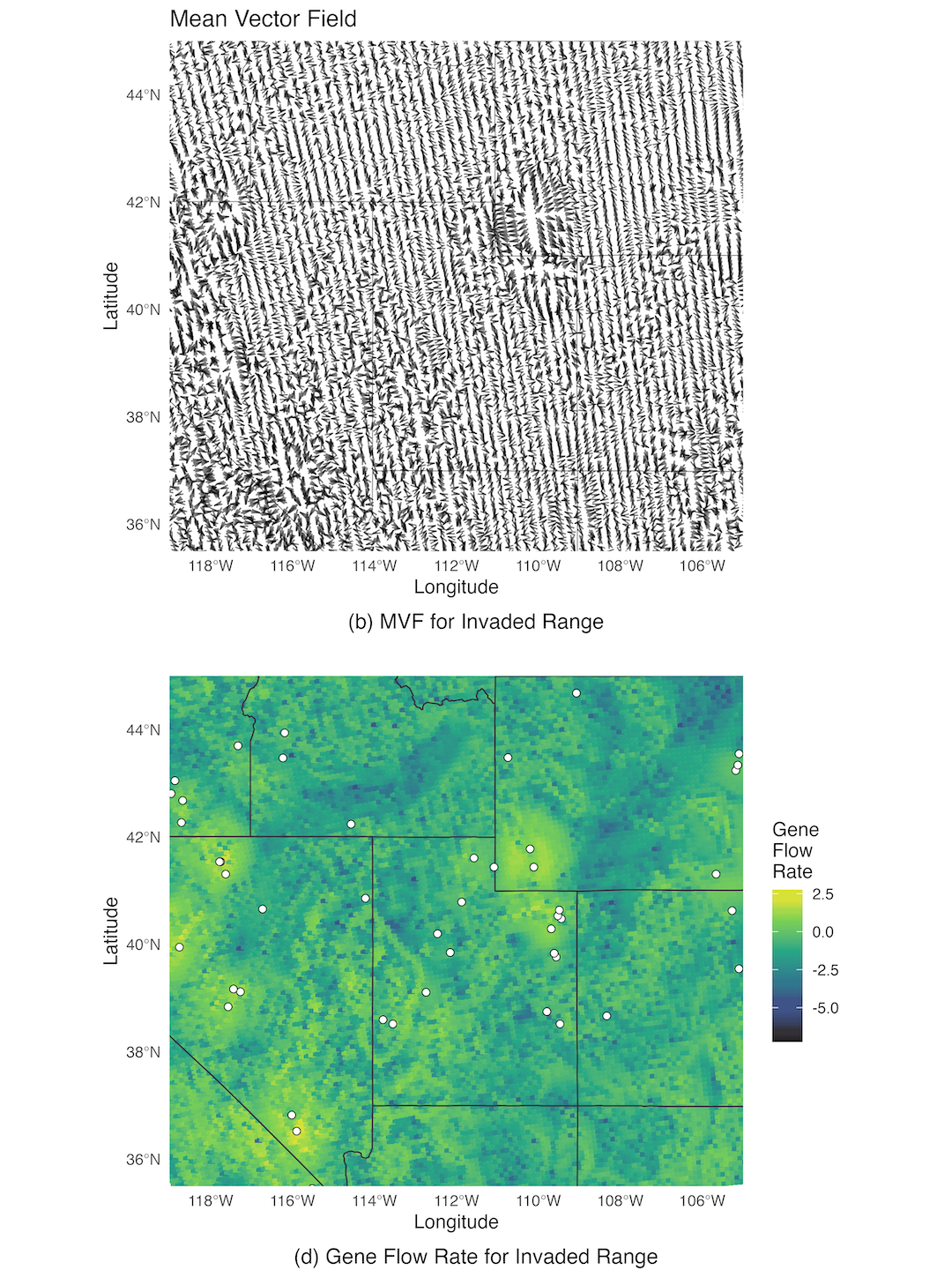}
  \end{subfigure}
  \caption{ \textit{Top}: Mean vector fields of gene flow. \textit{Bottom}: Maps of log-gradient magnitudes of gene flow (depicting gene flow rates) with white dots denoting observation locations. \textit{Left}: Inference in part of the native range (Germany, Poland, Czechia, Austria). \textit{Right}: Inference in part of the invaded range (Nevada, Idaho, Utah, Wyoming, Colorado).}
  \label{fig:mvf}
\end{figure}

Figure \ref{fig:mvf}(a-b) depicts the mean vector field of gene flow in portions of the native and invaded range. 
Arrow orientation indicates the dominant local direction of gene flow, and arrow length represents the relative magnitude, which we interpret as the strength of directional preference in dispersal.
We also quantified local gene flow rates as the logarithm of the spatial gradient magnitude of the predicted mean field (Figure \ref{fig:mvf}(c-d)). For each grid node, we computed the Euclidean norm of $u_g$ and $v_g$ as the gradient magnitude, which we then log-transformed to compress extreme values and emphasize fine-scale spatial variation. The gradient-based measure is particularly informative when gene flow is locally symmetric; opposing vector components may cancel in the mean field, whereas the gradient magnitude describes the strength of spatial variation in dispersal.

In the native range, the vector field is relatively smooth. Nearby arrows tend to align, suggesting relatively consistent dispersal directions across broad regions; however, there is localized heterogeneity in the mountainous regions because steep environmental gradients result in rapid shifts in flow direction. 
In contrast, the invaded range exhibits a more heterogeneous and less smooth vector field. While certain regions in the invaded range show coherent directional alignment, much of the Intermountain West is characterized by sharp environmental gradients and fragmented habitats. 
There appears to be more variation in directional preference in the invaded range, which suggests that gene flow in the invaded range is more strongly shaped by landscape heterogeneity and ongoing colonization processes. In summary, the relatively uniform vector field in the native range reflects the stability of an established species, whereas the more spatially variable vector field in the invaded range characterizes the dynamic, spatially heterogeneous pathways of an active invasion.

\section{Conclusion}
\label{sec:conclusion}

We proposed a Bayesian hierarchical dyadic model to infer gene flow in invasive species. Invasion dynamics are challenging to account for using existing landscape genomic models because invasion processes are often directional, inherently complex, and shaped by natural and anthropogenic pathways. We accounted for such challenges using a dyadic framework, which enabled us to model pairs of locations directly and to represent the asymmetric structure of invasive gene flow. We proposed dyadic SVCs to characterize spatially heterogeneous effects, and we introduced the concept of connectivity covariates to explicitly model gene flow related to discrete landscape features. 

We extended spatially varying coefficients to the dyadic space such that the effect of a landscape feature may vary across pairs of locations; our latent spatial factor specification represented those effects parsimoniously and avoided the prohibitive dimensionality of modeling each dyad separately. The DSVCs identified localized heterogeneity in environmental effects, and they retained IBD and IBR interpretability where those hypotheses held and recovered nonstationary invasion dynamics where they did not. The simulation study and the cheatgrass case study demonstrated that DSVCs significantly improved model performance under the presence of spatially heterogeneous effects.

Our second contribution was the development of connectivity covariates.
Dyadic analyses typically require covariates at the node level and cannot accommodate corridors or barriers between nodes. The connectivity covariates turn pathway information into dyadic covariates by condensing a high-dimensional spatial network into a low-dimensional summary and highlighting mechanisms of spread that environmental similarity alone cannot represent. They also provide mechanistic interpretability because the sign of a coefficient distinguishes a corridor from a barrier. For example, in conservation studies, the combination of DSVCs and connectivity covariates may be used to test the efficacy of wildlife corridors as a conservation practice \citep{lindenmayer1993ecological}.

The dominant computational cost of the sampler is the dense factorization of the $N \times N$ precision matrix for each latent factor at each iteration, which we describe in Appendix E and measure in Appendix H. A matrix-free implementation that replaces the factorization with preconditioned conjugate gradients that exploit the separable structure for the matrix-vector products is a natural route to scale for larger data sets. 
Another limitation is that the odd component of our model is only weakly identified when the dyads are unordered. Data in which dyads are genuinely directed, such as parentage assignments or time-stamped introductions, would identify the odd component directly and resolve asymmetry more finely than a symmetric dissimilarity.



Although motivated by invasive species, our framework is broadly applicable to any dyadic flow process that exhibits spatial heterogeneity. DSVCs and connectivity covariates provide general tools for uncovering nonstationary effects in the flow of processes. Potential applications extend to infectious disease dynamics, where transmission is mediated by heterogeneous contact networks; international trade, where flows are shaped by asymmetric policies and transportation routes; and ecological dispersal processes more broadly. By linking landscape features to the flow of processes, our model provides a flexible, interpretable, and computationally tractable framework for studying complex spatial dynamics across disciplines.  

\section*{Acknowledgments}
We thank the Bromecast research network for their data collection and contributions to this manuscript.

\section*{Funding}
This research was funded by the National Science Foundation: NSF 2222525, 1927177, 1927009, and 1927282.

\section*{Supplemental Materials}
\textit{All code and the derived data required to reproduce every table and figure can be found at \url{https://github.com/michaelschwob/nonstationary-dyadic-flow-models}.}

\bibliographystyle{plainnat}
\bibliography{bibliography}

\newpage 

\setcounter{figure}{0}
\setcounter{table}{0}
\renewcommand{\thefigure}{S\arabic{figure}}
\renewcommand{\thetable}{S\arabic{table}}
\renewcommand{\figurename}{Supplemental Figure}
\renewcommand{\tablename}{Supplemental Table}

\section*{Appendix A - Cheatgrass Data Preparation}

\subsection*{A.1 Genotype Data}

In what follows, we describe how we transformed single-nucleotide polymorphism (SNP) genotype data for cheatgrass into pairwise genetic distance matrices for the invaded and native ranges. 
The genotype data were organized as a table comprising locus metadata (e.g., chromosome, position, locus ID) in the first three columns followed by one column per individual genome sequence; we refer to the block of individual columns as the SNP matrix.
To ensure balanced genomic coverage, we subsetted the SNP matrix using a stratified random sample of loci by chromosome. In particular, we sampled 300 loci per chromosome for chr1-chr7 (i.e., $L=2100$ loci in total). 
The cost of forming the genetic distance matrix grows linearly in the number of loci and quadratically in the number of genotypes.
Stratification prevents an over-representation of genomic regions with many high-quality or high-minor-allele frequency loci, a bias that arises if filtering solely on call-rate or minor-allele frequency. 
By sampling within each chromosome stratum, we preserved broad, uniform coverage of the genome. 

After subsetting the loci in the invaded-range SNP matrix, we obtained the invaded-range genotype matrix $\mathbf{G}_{\text{inv}}\in\{0,1,2,\text{NA}\}^{L\times n_\text{inv}}$, where $L$ is the number of retained loci and $n_{\text{inv}}$ is the number of invaded-range individuals. 
Entries correspond to standard diploid dosage coding at each locus (i.e., 0/1/2 copies of the coded/reference allele, which was provided by the source SNP table); NA denotes a missing genotype call. We converted $\mathbf{G}_{\text{inv}}$ to a diploid \texttt{genind} object and computed the pairwise genetic distance matrix (GDM) in \texttt{R} using the \texttt{adegenet} package: $\texttt{adegenet::dist.gene}(\cdot, \texttt{method}=\texttt{"pairwise"})$, which yielded the symmetric GDM $\mathbf{D}_{\text{inv}}\in\mathbb{N}_0^{n_{\text{inv}}\times n_{\text{inv}}}$ with $D_{ij}=D_{ji}$ and $D_{ii}=0$. The GDM entries aggregate per-locus allele-sharing mismatches across the retained loci. For the native range, we followed the same steps to obtain $\mathbf{G}_{\text{native}}$ and $\mathbf{D}_{\text{native}}$ from the native-range SNP matrix. 

We made several implementation decisions to support comparability across ranges and downstream integration. First, we assumed diploidy throughout (\texttt{ploidy}$=2$ in \texttt{df2genind}).
Second, while the number of retained loci $L$ can differ between ranges, we used the same stratified design (300 loci per chromosome across chr1-chr7) for both ranges, so $L$ was identical in both ranges; this standardized scale and variance of pairwise distances across the ranges. Third, \texttt{dist.gene} computes distances using pairwise deletion (i.e., for each pair of individuals, it aggregates only over loci observed in both), which mitigates bias from unequal call rates without imputing genotypes. Finally, we preserved site/genotype identifiers as row and column names, enabling direct spatial joins and alignment with environmental covariates. These steps yielded consistent, allele-sharing-based GDMs for both native and invaded ranges that were suitable for mapping, downstream modeling, and cross-range comparisons.


\subsection*{A.2 Alternative Measures of Genetic Dissimilarity}
We define $d_{ij}$ as the number of discordant loci because cheatgrass is highly selfing and heterozygotes are rare; however, in diploid species with frequent heterozygosity, $d_{ij}$ may be computed from allele-level mismatch scores (e.g., 0, 1, or 2 mismatches per locus). 
Equivalently, the discordant counts $d_{ij}$ may be modeled directly with $d_{ij}\sim\text{Binomial}(M_{ij},\text{logit}^{-1}(z_{ij}))$, where $z_{ij}$ is a linear predictor that is naturally weighted by $M_{ij}$.

\section*{Appendix B - Dyadic Covariance Structure} 
We form the separable covariance as the Kronecker structure $\mathbf{K}^s(\phi_q)\otimes\mathbf{K}^d(\phi_q)$, where $\mathbf{K}^s(\phi_q)\in\mathbb{R}^{n\times n}$ and $\mathbf{K}^d(\phi_q)\in\mathbb{R}^{n\times n}$ are spatial covariance matrices defined over the first and second members of a dyad, respectively; this accommodates potentially different spatial processes at the two positions. 
However, separate node-level covariance structures increase the number of spatial range parameters, which are difficult to estimate \citep{zhang2004inconsistent}.  
Given the quadratic growth of dyads (i.e., $N={n \choose 2}$), such flexibility can lead to overfitting.
Therefore, we assume symmetry in the node-level covariance matrices (i.e., $\mathbf{K}=\mathbf{K}^s=\mathbf{K}^d$); this assumption is appropriate in many ecological and invasion contexts, where the same spatial processes govern both members of a dyad (e.g., geographic distance, habitat continuity). 

To induce symmetry in the node-level covariance matrix $\mathbf{K}(\phi_q)$, we let $\mathbf{H}\in\{0,1\}^{n^2\times n^2}$ denote the commutation matrix that satisfies $\mathbf{H}\text{vec}(\mathbf{A})=\text{vec}(\mathbf{A}')$ and $\mathbf{H}(\mathbf{A}\otimes \mathbf{B})\mathbf{H}=\mathbf{B}\otimes \mathbf{A}$ for any $n \times n$ matrices $\mathbf{A}$ and $\mathbf{B}$, where $\text{vec}(\cdot)$ stacks the columns of a matrix to form a vector.  
Then, we construct the symmetrized Kronecker structure over the full node-node space: 
$$\mathbf{K}^*_q = (\mathbf{I}+\mathbf{H})(\mathbf{K}(\phi_q)\otimes \mathbf{K}(\phi_q))(\mathbf{I}+\mathbf{H})'\in\mathbb{R}^{n^2\times n^2},$$ 
where $\mathbf{I}$ is the identity matrix. We restrict the covariance matrix to the dyadic index set $\{i<j\}$ using a fixed mapping matrix $\mathbf{E}\in\mathbb{R}^{N\times n^2}$:
\begin{equation}\label{eq:kron}
    \boldsymbol{\Sigma}_q=\mathbf{E}\mathbf{K}^*_q\mathbf{E}'\in\mathbb{R}^{N\times N},
\end{equation}
where each row of $\mathbf{E}$ selects the entry corresponding to the dyad $(i,j)$ with $i<j$. By restricting the node-node kernel $\mathbf{K}^*_q$ to the set $\{i<j\}$, we enforce a consistent orientation and analyze each unordered pair of nodes only once.

\section*{Appendix C - Sensitivity Analyses}
 
\subsection*{C.1 The Spatial Scale of the Connectivity Covariates}
 
Each column of the pathway distance data is the distance from a genotype to one named feature (e.g., a specific interstate, a specific creek). Therefore, a decay parameter set from the median over all such distances describes the distance to a randomly chosen feature anywhere in the study region rather than the distance at which a pathway plausibly affects dispersal. In our case study, the median corresponds to a half-decay distance of 848 km.
 
We report the spatial scale of the pathway classes in our case study in Supplemental Table \ref{tab:g1} with the participation ratio $(\sum_j v_{ij,c})^2/\sum_j v_{ij,c}^2$, which counts the features that effectively contribute to the closeness score of node $i$ in class $c$. A value near $n_c$ means every feature contributes equally, and the closeness score records position rather than proximity to any particular pathway. At a half-decay distance of 848 km, 674 of the 745 river features contributed to the median node. In contrast, when we set the decay as the median nearest-feature distance, only 2 features contributed to the median node. In the fits that estimate $\tau_c$, we centered its prior on the median nearest-feature distance within class $c$ to obtain a higher resolution of the pathway effects.
 
\begin{table}[h]
\caption{Spatial scale of the three pathway classes. The half-decay distance is the distance at which a feature contributes half of its maximum closeness. Distances enter the closeness score in units of 10 km, so this distance is $10\tau_c\log(2)$ km. ``Effective features'' is the median over nodes of the participation ratio.}
\begin{tabular}{l|rrrrrr}
\label{tab:g1}
Class & $n_c$ & Nodes within 25 km & \multicolumn{2}{c}{Half-decay (km)} & \multicolumn{2}{c}{Effective features} \\
      &       &                    & all-pairs & nearest & all-pairs & nearest \\ \hline
railroad    & 2   & 11 & 848 & 222.3 & 1.9   & 1.39 \\
highway     & 20  & 47 & 848 & 19.3  & 18.3  & 1.15 \\
river/creek & 745 & 77 & 848 & 5.0   & 674.3 & 2.44
\end{tabular}
\end{table}

Supplemental Table \ref{tab:g1} also identifies a limit of the pathway data. The railroad class comprises two features, and only 11 of the 86 sampled genotypes lie within 25 km of either. Therefore, a small decay yields a covariate that is zero for almost every node, and a large decay yields a smooth function of position; neither measures proximity to a railroad. The highway class has a different problem with the same consequence; its coefficient is not stable across decays, and the decay itself is determined. As a result, the sign reverses between the fixed-decay and estimated-decay approaches while both credible intervals exclude zero. The river class is the only class whose sign is stable across both approaches.
 
\subsection*{C.2 Sensitivity of the Connectivity Coefficients to the Decay}
 
We quantified how much the connectivity covariates vary with decay. Supplemental Table \ref{tab:g2} reports Pearson correlations between the standardized connectivity covariates computed at $0.25\tau^0_c$, $0.5\tau^0_c$, $2\tau^0_c$, and $4\tau^0_c$ and those computed at $\tau^0_c$ itself.
Because the sensitivity of the decay is not negligible, we let the data inform $\{\tau_c\}$.
 
\begin{table}[h]
\caption{Pearson correlation between the standardized connectivity covariates at each scaled decay and at the decay itself.}
\begin{tabular}{l|rrrr}
\label{tab:g2}
Multiplier & 0.25 & 0.5 & 2 & 4 \\ \hline
railroad     & 0.855 & 0.973 & 0.991 & 0.979 \\
highway     & 0.737 & 0.924 & 0.976 & 0.949 \\
river/creek & 0.781 & 0.948 & 0.976 & 0.942
\end{tabular}
\end{table}

To further investigate the sensitivity of $\{\tau_c\}$, we fit the connectivity-only model at 0.5, 1 and 2 times the all-pairs decay (Supplemental Table \ref{tab:g3}). The river coefficient was negative at every decay with a credible interval excluding zero. The highway coefficient was positive at every decay, although its magnitude tripled and the interval excluded zero only for the larger decays. The railroad coefficient reversed sign.
Between the fixed-decay and the estimated-decay fits, the highway coefficient moves from 0.047 $[0.019,0.078]$ to $-0.083$ $[-0.114,-0.049]$, so both intervals exclude zero but in opposite directions. Therefore, we report the connectivity covariates as contributing to fit without attributing an effect to any single pathway class. Across all three fits, the dyadic variance $\sigma^2$ moved by roughly 0.3\%. Therefore, the decay does not change how well the model describes the data, though it does change which pathway class receives the credit.
 
\begin{table}[h]
\caption{Connectivity coefficients from the connectivity-only invaded-range fit at three fixed decays with 95\% credible intervals.}
\begin{tabular}{l|rrr}
\label{tab:g3}
 & $0.5\tau$ & $\tau$ & $2\tau$ \\ \hline
railroad    & $+0.019$ [$-0.010$, $0.048$] & $-0.019$ [$-0.049$, $0.010$] & $-0.038$ [$-0.066$, $-0.010$] \\
highway     & $+0.022$ [$-0.011$, $0.056$] & $+0.047$ [$0.019$, $0.078$]  & $+0.068$ [$0.039$, $0.096$] \\
river/creek & $-0.123$ [$-0.151$, $-0.096$] & $-0.111$ [$-0.135$, $-0.087$] & $-0.095$ [$-0.118$, $-0.072$] \\
$\sigma^2$  & 0.3573 $[0.3414, 0.3740]$ & 0.3568 $[0.3407, 0.3736]$ & 0.3562 $[0.3404, 0.3732]$
\end{tabular}
\end{table}
 
Estimating $\{\tau_c\}$ as described in Section 4.2 replaces the sensitivity with a posterior. Supplemental Table \ref{tab:g4} reports the posterior half-decay distances from (i2) and (i4). Model (i2) determines all three classes with effective sample sizes between 391 and 619 and $\hat{R}$ no larger than 1.007. The full fit fails to determine every pathway class; its railroad interval spans an order of magnitude, its effective sample sizes fall as low as 20, and the chain for the river decay did not converge. Therefore, the decay is identifiable in the absence of a nonstationary coefficient surface and is not identifiable in its presence; this is the same conclusion that the coefficients support in Section 7.2. Estimating the decays did not improve the fit of either model: WAIC was 6675.8 for (i2) and 4929.5 for (i4) with the decays estimated, against 6674.5 and 4914.7 with them fixed (Table 2).
 
\begin{table}[h]
\caption{Posterior half-decay distances in kilometers with 95\% credible intervals from (i2) and (i4). Both fits estimate $\{\tau_c\}$.}
\begin{tabular}{l|rr}
\label{tab:g4}
Class & (i2) connectivity only & (i4) full model \\ \hline
railroad    & 98.9 [70.6, 132.5]    & 930.1 [254.0, 1727.9] \\
highway     & 198.0 [128.3, 266.6]  & 44.1 [19.1, 56.4] \\
river/creek & 15.1 [9.1, 22.7]      & 44.6 [1.3, 74.7]
\end{tabular}
\end{table}
 
\subsection*{C.3 The Number of Radial Basis Function Centers}
 
In Section 7, we used five radial basis function centers. To justify that choice, we performed five-fold cross-validation of a ridge regression of the dyadic response on the radial basis expansion alone, where the centers and the bandwidth were chosen within each training fold.
The cross-validated $R^2$ rises monotonically with the number of centers: $-0.001$ at 2 centers, 0.027 at 3, 0.041 at 5, 0.054 at 8, 0.066 at 12, and 0.069 at 20.
We note that the criterion omits the DSVC layer, so it bounds what a richer stationary basis can provide. To measure that bound directly, we fit (i2) at 5, 12 and 20 centers, and $R^2$ on the fitted dyads rose from 0.175 to 0.245 to 0.260. In contrast, the full model (i4) at five centers reaches $R^2=0.816$. 
Therefore, a richer stationary basis recovers a small fraction of what the DSVCs recover.

We repeated the cross-validation across kernel bandwidths at 0.5$\times$, 1$\times$, and 2$\times$ the median inter-center distance. The criterion was close to flat in the bandwidth. At twenty centers, it takes the values 0.06, 0.069, and 0.069, respectively, and at five centers, it takes the values 0.021, 0.041, and 0.050, respectively. Therefore, the number of centers matters more than the bandwidth, though neither choice moves the criterion by as much as adding the coefficient surface.
 
Regardless, we used five centers throughout our analyses.
The largest basis that we tested raised $R^2$ on the fitted dyads from 0.175 to 0.260 without a surface, whereas adding the surface at five centers raised it to 0.816. The stationary basis and the coefficient surface are substitutes, and the surface accounts for most of the structure that either can represent, so the size of the basis is not a consequential choice in our analyses. Additionally, setting the latent factor dimension to $Q=p+C$ makes the sampling cost linear in the size of the basis. A larger basis would cost runtime and additional spatial range parameters that we argue are not identifiable in Section 4.1.
 
\subsection*{C.4 Effective Rank of the Fitted Coefficient Surface}

We did not select the latent factor dimension $Q$. Instead, we set $Q=p+C$ and let the global-local shrinkage prior on $\mathbf{C}$ decide how much of that space to use. Here, we investigate how much rank the fitted surface actually used in our case study and what the cost is of setting $Q$ too small or too large.
 
Section 4.1 showed that $\boldsymbol\Delta=\mathbf{WC}'$ was identified only up to an invertible $Q\times Q$ transformation, so individual latent factors and their corresponding $\phi_q$ were not identifiable. 
However, the singular values of the posterior mean of $\boldsymbol\Delta$ were invariant to the transformation and gave an effective rank for the fitted surface. In the full invaded-range fit with $Q=8$, the first two factors accounted for 94.4\% of the surface and the first three for 99.0\%. In the native fit with DSVCs with $Q=5$, the first two factors accounted for 99.1\%. Thus, the shrinkage prior on $\mathbf{C}$ selected an effective rank well below $Q$ in both ranges, and setting $Q=p+C$ did not impose the dimension it appeared to impose.
 
To determine the cost of setting $Q$ too small or too large, we fit the full model under (s1) with $Q\in\{2,6,12\}$ against a true rank of six, and we fit the models containing connectivity covariates under (s1) with the true decay multiplied by 0.5, 1, and 2.
We report the sensitivity of such choices in Supplemental Tables \ref{tab:sim-Q} and \ref{tab:sim-tau}.
 
\begin{table}
\caption{Sensitivity to the number of latent spatial factors. Each cell is the mean over 25 replicates with the Monte Carlo error in parentheses. Data are generated with true rank 6. $Q = p+C=6$ is the default, under which the global-local shrinkage prior on $\mathbf{C}$ selects the effective rank.}
\label{tab:sim-Q}
\centering\small\begin{tabular}{lrrrrr}
\toprule
Fitted $Q$ & $\boldsymbol{\beta}$ RMSE & $\boldsymbol{\Delta}$ RMSE & $\boldsymbol{\Delta}$ coverage & CRPS & it/s \\
\midrule
2 & 0.043 (0.006) & 0.149 (0.011) & 0.86 & 0.205 (0.007) & 32.75 \\
6 & 0.040 (0.004) & 0.141 (0.010) & 0.96 & 0.188 (0.004) & 17.57 \\
12 & 0.039 (0.004) & 0.142 (0.010) & 0.97 & 0.183 (0.004) & 10.73 \\
\bottomrule\end{tabular}\end{table}
 
\begin{table}
\caption{Sensitivity of the full model to the connectivity decay $\tau_c$, reported as a multiple of the value used in the simulation design. Each cell is the mean over 25 replicates with the Monte Carlo error in parentheses.}
\label{tab:sim-tau}
\centering\small\begin{tabular}{lrrrr}
\toprule
$\tau$ multiplier & $\boldsymbol{\beta}$ RMSE & $\beta_5$ bias & $\beta_6$ bias & CRPS \\
\midrule
0.50 & 0.054 (0.006) & -0.006 & 0.033 & 0.189 (0.004) \\
1.00 & 0.040 (0.004) & -0.014 & 0.013 & 0.188 (0.004) \\
2.00 & 0.031 (0.003) & -0.005 & 0.005 & 0.187 (0.004) \\
\bottomrule\end{tabular}\end{table}
 
\subsection*{C.5 Prior Sensitivity of the Latent Spatial Variance}
 
The node-level latent field has $n-1$ effective degrees of freedom. As a result,
$\sigma^2_\eta$ is identified by little information. We fit (n1) under $\sigma^2_\eta\sim\text{InvGamma}(2,1)$ against the
$\text{InvGamma}(0.1,0.1)$ specification of the reported analysis and report the comparison in Supplemental Table \ref{tab:g6}. We chose $\text{InvGamma}(2,1)$ as the comparison because the reported $\text{InvGamma}(0.1,0.1)$ prior is nearly improper and does not have a finite mean, whereas $\text{InvGamma}(2,1)$ is proper and informative with a mean of one. Therefore, if the posterior shifts very little between the two priors, we can conclude that there is little prior sensitivity.
 
\begin{table}[h]
\caption{Posterior means and 95\% credible intervals for the native standard fit under two different priors for $\sigma^2_\eta$.}
\label{tab:g6}
\centering\small
\begin{tabular}{l|rr}
Parameter & $\text{InvGamma}(0.1,0.1)$ & $\text{InvGamma}(2,1)$ \\ \hline
$\alpha$          & $-0.7545$ [$-0.7774$, $-0.7310$] & $-0.7540$ [$-0.7771$, $-0.7308$] \\
$\beta_1$         & $0.1566$ [$0.0957$, $0.2175$]    & $0.1565$ [$0.0933$, $0.2208$] \\
$\beta_2$         & $0.2143$ [$0.1510$, $0.2768$]    & $0.2176$ [$0.1553$, $0.2800$] \\
$\beta_3$         & $-0.0123$ [$-0.1105$, $0.0838$]  & $-0.0140$ [$-0.1099$, $0.0817$] \\
$\beta_4$         & $-0.5176$ [$-0.5588$, $-0.4766$] & $-0.5167$ [$-0.5559$, $-0.4776$] \\
$\beta_5$         & $-0.1156$ [$-0.1710$, $-0.0587$] & $-0.1169$ [$-0.1715$, $-0.0594$] \\
$\sigma^2$        & $0.0709$ [$0.0683$, $0.0736$]    & $0.0709$ [$0.0682$, $0.0738$] \\
$\sigma^2_\eta$   & $0.0200$ [$0.0073$, $0.0533$]    & $0.1732$ [$0.0822$, $0.3564$] \\
$\phi_\eta$       & $9.78$ [$1.77$, $29.64$]         & $95.00$ [$38.53$, $205.68$]
\end{tabular}
\end{table}
 
Only $\sigma^2_\eta$ and $\phi_\eta$ were sensitive to the prior for $\sigma^2_\eta$.
The posterior mean of $\sigma^2_\eta$ moved by a factor of 8.7, and the two
credible intervals were disjoint. The spatial range moved by a factor of
9.7, and its intervals were also disjoint. In contrast, the mean structure and the dyadic variance remained unchanged.
Such a pattern follows from the geometry of the latent field. The variance and the
range of a Gaussian process observed at 105 locations were weakly separated
because the data identified their combination better than either alone. Thus, a prior that pushed $\sigma^2_\eta$ upward was accommodated by lengthening the range
rather than by changing the mean structure. The effective sample size for $\sigma^2_\eta$ rose from 127 to 308 under the tighter prior, which was the prior
supplying identification that the data did not.
Under
the $\text{InvGamma}(2,1)$ prior, the posterior mean of $\phi_\eta$ was 95.0 against a prior
median of 9.3, which was 1.55 prior standard deviations above the prior median on the log scale and was in the upper 6\% of its own prior. Therefore, we retain the $\text{InvGamma}(0.1,0.1)$ specification because it is the weakly informative default and nothing we interpret depends on the choice.
 
\subsection*{C.6 A Scale Invariance in the Latent Factorization}
 
Dividing the $q$th column of $\mathbf{W}$ by a constant and multiplying the $q$th column of $\mathbf{C}$ by that same constant leaves $\boldsymbol\Delta$ unchanged, so the likelihood is flat along that direction. The prior on $\mathbf{w}_q$ is densest at zero, so shrinking a column of factor scores toward zero is not penalized. In fact, the heavy tails of the global-local prior on $\mathbf{c}_q$ mean that concentrating growth in the loadings is barely penalized. As a result, the posterior can move along the ridge without the density falling. Therefore, the posterior admits a ridge along which a factor collapses while the fitted surface remains fixed.
 
Model (n2) traversed that ridge, whereas (i3) did not. Supplemental Table \ref{tab:g5} reports the within-draw variance of each column of $\mathbf{W}$, which the prior sets to one, along with the norm of the corresponding column of $\mathbf{C}$. In (i3), every column stays near the prior scale. In (n2), the fifth column has a factor-score variance of 0.0002 and a loading norm of 248; the product of its standard deviation and its loading norm, which is the identified quantity, is the largest of the five.
 
\begin{table}[h]
\caption{Position of each latent factor along the scale ridge. The prior sets the variance of $\mathbf{w}_q$ to one. The identified quantity is the product of the factor-score standard deviation and the loading norm along with their 95\% credible interval; the two coordinates separately are not identified.}
\begin{tabular}{l|rrr|rrr}
\label{tab:g5}
 & \multicolumn{3}{c|}{(i3) invaded, DSVCs} & \multicolumn{3}{c}{(n2) native, DSVCs} \\
Factor & $\mathrm{var}(\mathbf{w}_q)$ & $\|\mathbf{c}_q\|$ & product & $\mathrm{var}(\mathbf{w}_q)$ & $\|\mathbf{c}_q\|$ & product \\ \hline
1 & 0.714 & 0.89 & 0.58 [0.01, 1.87] & 0.628  & 1.65   & 1.30 [0.99, 1.88] \\
2 & 0.824 & 0.28 & 0.23 [0.01, 0.67] & 0.490  & 0.50   & 0.33 [0.14, 0.54] \\
3 & 0.829 & 0.29 & 0.25 [0.01, 0.80] & 0.940  & 0.40   & 0.39 [0.34, 0.45] \\
4 & 0.963 & 1.52 & 1.49 [1.32, 1.68] & 0.864  & 0.36   & 0.33 [0.21, 0.48] \\
5 & 0.300 & 3.74 & 1.38 [0.03, 2.07] & 0.0002 & 247.86 & 3.45 [2.79, 4.21]
\end{tabular}
\end{table}
 
The reason behind such a phenomenon is how we propose in the MCMC algorithm. The whitened joint random-walk proposal for $(\phi_q,\mathbf{w}_q)$ in Appendix E moves $\boldsymbol\Delta$ by an amount scaled by $\|\mathbf{c}_q\|$, so a factor far along the ridge takes an effective step in $\boldsymbol\Delta$ two orders of magnitude larger than intended. Acceptance for the range parameters in (n2) fell to between 0.04\% and 6\% against 13\% to 62\% in (i3). Additionally, $\phi_5$ accepted no proposal across the 2,500 retained draws, and the chains for $\alpha$ and $\boldsymbol\beta$ inherit that behavior through their dependence on $\boldsymbol\Delta$.
 
We could close the ridge without changing the posterior by renormalizing each column of $\mathbf{W}$ to unit variance at the end of every iteration and multiplying the corresponding column of $\mathbf{C}$ by the removed constant. However, we did not make such modifications because the fitted surface was unaffected by the position of a factor along the ridge. Rather, we computed the posterior mean of $\boldsymbol\Delta$ on four consecutive blocks of the retained draws for (n2) and for (i3). In both fits, the disagreement between the first block and the last was no larger than the disagreement between adjacent blocks.


\section*{Appendix D - Bayesian Hierarchical Model}\label{app:model}
\begin{align*}
    \widetilde{y}_{ij} &\sim \text{N}\left(\alpha +\mathbf{z}_{ij}'(\boldsymbol{\beta}+\boldsymbol{\delta}_{ij})+\widetilde{\eta}_{ij},\sigma^2\right),\\
    \alpha &\sim \text{N}(0, 10^6),\\
    \boldsymbol{\beta} &\sim \text{N}(\boldsymbol{0},10^6\mathbf{I}),\\
    \boldsymbol{\eta} &\sim \text{N}(\boldsymbol{0},\sigma^2_\eta\mathbf{R}(\phi_\eta)),\\
    \sigma^2,\sigma^2_\eta &\sim \text{IG}(0.1,0.1),\\
     \boldsymbol{\Delta} &= \mathbf{WC}',\\
    \mathbf{w}_q &\sim \text{N}(\boldsymbol{0},\boldsymbol\Sigma_q(\phi_q)),\\
    \mathbf{c}_q &\sim \text{N}(\boldsymbol{0},\mathbf{D}_q),\\
    \log(\phi_\eta),\log(\phi_q) &\sim \text{N}(\mu_{\log\phi},2.25),\\
    \mathbf{D}_q &= \text{diag}\{\lambda^2_{lq}\xi^2_q\},\\
    \lambda_{lq},\xi_q &\sim \text{C}^+(0,1),\\
    \log(\tau_c) &\sim \text{N}(\log\tau^0_c,(\log 4/2)^2),
\end{align*}
where $\mathbf{z}_{ij}=[\widetilde{\mathbf{x}}'_{ij} \;\boldsymbol{\kappa}_{ij}']'$, each row in $\boldsymbol\Delta\in\mathbb{R}^{N\times (p+C)}$ corresponds to $\boldsymbol\delta_{ij}'$, $\mu_{\log\phi}=\log(\text{median}(\mathbf{D}_{\text{euc}}))$, $\mathbf{D}_{\text{euc}}$ is the Euclidean distance matrix between observations, and $\mathbf{w}_q$ and $\mathbf{c}_q$ are the $q$th columns of $\mathbf{W}$ and $\mathbf{C}$, respectively. The connectivity covariates $\boldsymbol{\kappa}_{ij}$ depend on the decay parameters $\{\tau_c\}^C_{c=1}$ through Section 4.2, and $\tau^0_c$ is the median distance from a node to the nearest feature in class $c$ divided by $\log2$.

\section*{Appendix E - Derivations of the MCMC Updates}

\subsection*{E.1 Updating $\boldsymbol{\eta}$ in the Reduced Subspace}
To avoid confounding with the intercept, we projected $\boldsymbol{\eta}$ into the $(n-1)$-dimensional subspace orthogonal to the constant vector $\mathbf{1}_n$.
In particular, we set $\boldsymbol{\eta}=\mathbf{U}\boldsymbol{\gamma}$, where $\mathbf{U}$ is an $n\times(n-1)$ orthonormal basis satisfying $\mathbf{U}'\mathbf{1}=\mathbf{0}$ and $\boldsymbol\gamma\sim\text{N}(\boldsymbol0,\sigma^2_\eta\mathbf{U}'\mathbf{R}_\eta(\phi_\eta)\mathbf{U})$. Updates were then performed in the $\mathbf{U}$-subspace with the conjugate posterior $\boldsymbol\gamma \sim \text{N}(\boldsymbol\mu_\gamma,\boldsymbol\Sigma_\gamma)$, where $$\boldsymbol\Sigma_\gamma=\left(\frac{(\mathbf{MU})'\mathbf{MU}}{\sigma^2}+\frac{(\mathbf{U}'\mathbf{R}_\eta(\phi_\eta)\mathbf{U})^{-1}}{\sigma^2_\eta}\right)^{-1},\quad \boldsymbol\mu_\gamma=\boldsymbol\Sigma_\gamma\frac{(\mathbf{MU})'\widetilde{\mathbf{y}}}{\sigma^2},$$
$\mathbf{M}$ is the $N \times n$ dyadic mapping matrix for $\boldsymbol\eta$, and $\widetilde{\mathbf{y}}$ is the vector comprising all dyadic responses.
After drawing $\boldsymbol{\gamma}$, we reconstructed $\boldsymbol{\eta}=\mathbf{U}\boldsymbol{\gamma}$. We updated $\boldsymbol\eta$ in the reduced subspace because it enforced identifiability by removing the uninformative constant that would otherwise be confounded with the intercept. Additionally, working in the $\mathbf{U}$-subspace improved numerical stability because $\mathbf{U}'\mathbf{R}_\eta(\phi_\eta)\mathbf{U}$ is positive-definite, whereas the full covariance $\mathbf{R}_\eta(\phi_\eta)$ becomes nearly singular at large spatial ranges.

\subsection*{E.2 Sampler Settings}
For the slice sampler on $\log\phi_\eta$ and $\log\phi_q$, we initialized the bracket width as $w=0.5$ and performed up to 50 step-outs after a short burn-in period \citep{neal2003slice}.
We used random-walk proposals in the slice sampler with reflective bounds tied to the observed distance range and the $(\mu_{\log\phi}\pm 3\sigma_{\log\phi})$ prior window at a proposal standard deviation of $\sigma_{\text{rw}}=0.15(\log\phi_{\max}-\log\phi_{\min})$, which yielded acceptance rates near 0.30.  
Then, proposals were drawn uniformly from this interval until one fell within the slice. 
This procedure avoided manual tuning of proposal variances and ensured efficient exploration in the parameter space. We considered informative hyperparameters for the $\phi_\eta$ prior to regularize estimation \citep{zhang2004inconsistent}: $\mu_{\log\phi}=\log(\text{median}(\mathbf{D}_{\text{euc}}))$ and $\sigma^2_{\log\phi}=2.25$, where $\mathbf{D}_{\text{euc}}$ is the pairwise Euclidean distance matrix of the observed nodes.

Appendix D gives the priors for $\mathbf{w}_q$ and $\log\phi_q$,
where $\boldsymbol\Sigma_q(\phi_q)$ is the dyadic covariance matrix constructed from the node-level spatial kernel at range $\phi_q$. Because $\phi_q$ and $\mathbf{w}_q$ are strongly correlated, updating them separately can yield poor mixing. To stabilize MCMC sampling, we alternated between the two complementary update schemes that follow.

\subsection*{E.3 Whitened Joint Random-Walk Proposal for $(\phi_q,\mathbf{w}_q)$}

We expressed $\mathbf{w}_q$ in whitened coordinates as $\mathbf{v}_q=\mathbf{L}_q^{-1}\mathbf{w}_q$, where $\mathbf{L}_q$ is the Cholesky factor of $\boldsymbol\Sigma_q(\phi_q)$. Then, we proposed a new value for the spatial range parameter with $\log\phi_q^{(*)}\sim\text{N}(\log\phi_q^{(k-1)},\sigma^2_{rw})$. We sampled $\log\phi_q^{(*)}$ until we obtained a proposal in the interval $[\log\phi_{\min},\log\phi_{\max}]$.
Conditional on $\phi_q^{(*)}$, we mapped back to the latent factor with
$$\mathbf{w}_q^{(*)} \;=\; \mathbf{L}_q^{(*)}\,\mathbf{v}_q$$
to ensure that the proposed $(\phi_q^{(*)},\mathbf{w}_q^{(*)})$ was deterministically consistent, where $\mathbf{L}_q^{(*)}$ is the Cholesky factor of the proposed dyadic covariance $\boldsymbol\Sigma_q(\phi_q^{(*)})$. This joint proposal was accepted or rejected with the usual Metropolis-Hastings ratio. In practice, we only attempted this move when the factor $\mathbf{w}_q$ was non-negligible (i.e., when $\|\mathbf{c}_{q}\|$ and $\mathrm{var}(\mathbf{Zc}_{q})$ exceeded small thresholds), thereby avoiding expensive factorizations for weakly identified latent factors.

\subsection*{E.4 Conditional Gaussian Update for $[\mathbf{w}_q\mid\phi_q]$}

Given $\phi_q$, the full-conditional distribution for the $q$th latent dyadic spatial factor was
$$\mathbf{w}_{q}\mid \phi_q \;\sim\; \text{N}\!\big(\boldsymbol{\mu}_{q},\boldsymbol{\Sigma}_{q}^*\big),$$
where
$$\boldsymbol{\Sigma}_{q}^* = \Big(\boldsymbol\Sigma_q(\phi_q)^{-1}+\frac{1}{\sigma^2}\text{diag}(\mathbf{s}_q^{\circ 2})\Big)^{-1}, 
\qquad
\boldsymbol{\mu}_{q} = \boldsymbol{\Sigma}_{q}^*\,\frac{1}{\sigma^2}\text{diag}(\mathbf{s}_q)\,\mathbf{r}_{-q},$$

$$\mathbf{s}_q \equiv \mathbf{Z}\mathbf{c}_{q}, \qquad \mathbf{r}_{-q} \equiv \mathbf{\widetilde{y}}-\alpha\mathbf{1}_N-\mathbf{Z}\boldsymbol{\beta}-\mathbf{M}\boldsymbol{\eta}
-\sum_{q'\neq q}\big(\mathbf{s}_{q'}\odot \mathbf{w}_{q'}\big),$$
and the superscript $({\circ 2})$ denotes element-wise squares, $\mathbf{Z}$ is the design matrix comprising pairwise differenced environmental covariates and connectivity covariates for all observations, and $\odot$ denotes the Hadamard product.
The partial residual $\mathbf{r}_{-q}$ removes the column contribution to the likelihood such that it is the signal that the $q$th latent factor is left to explain.
If $\mathbf{s}_q\equiv \boldsymbol0$, then $\boldsymbol{\Sigma}_{q}^*=\boldsymbol\Sigma_q(\phi_q)$, $\boldsymbol{\mu}_{q}=\boldsymbol0$, and the posterior equals the prior. 

We note that the separable structure does not survive in this full conditional update. The precision of $[\mathbf{w}_q\mid\cdot]$ is the sum of the inverse dyadic covariance and a diagonal matrix, and a sum of this form does not have a Kronecker factorization. Therefore, we form the $N\times N$ precision matrix and factorize it densely at $\mathcal{O}(N^3)$ cost for each factor at each iteration. However, the separable structure is still useful because we use it to build $\boldsymbol{\Sigma}_q$ from an $n\times n$ node-level kernel rather than assembling an $N\times N$ matrix directly; it also ensures that the result is positive-definite. We explore the cost of the factorization and its scaling in Appendix H.

\subsection*{E.5 Cost of the Decay Update}

Each Metropolis proposal for $\tau_c$ required recomputing the closeness scores $\mathbf{V}_c$, the shared-segment scores $\mathbf{S}^*_c$, and the corresponding column of the dyadic design matrix at a cost of $\mathcal{O}(n^2n_c)$ operations for a class comprising $n_c$ features. A change in $\tau_c$ affected one column of the design, and the residual was updated in a single pass over the $N$ dyads rather than by recomputing the mean structure. In the full model, the resulting cost was a fraction of the $Q$ dense Cholesky factorizations that the latent factor update required. Over the 50,000 iterations, the decay updates took 6.2 seconds in (i2) and 2.6 seconds in (i4). The runtimes in Table 2 of the manuscript are for the fixed-decay fits, which skip this update.

\section*{Appendix F - Simulation Study Design and Model Comparison}

\subsection*{F.1 Construction of the Data-Generating Processes}

Section 6 states which features of the model each of (s1), (s2), and (s3) misspecifies. The node-level covariates were drawn as $\mathbf{x}_i\sim\text{N}(\boldsymbol{0},25\mathbf{I}_4)$ for $i=1,\dots,n$; we standardized each column and took pairwise differences over $\{i<j\}$ to obtain $\widetilde{\mathbf{x}}_{ij}$. The horizontal barrier was centered at $y=0.5$ and the vertical corridor was centered at $x=0.5$, each forming a class with one feature. Then, we computed $\boldsymbol{\kappa}_{ij}\in\mathbb{R}^2$ as in Section 4.2 with $\tau_1=\tau_2=0.07$ such that $\mathbf{z}_{ij}=(\widetilde{\mathbf{x}}_{ij}',\boldsymbol{\kappa}'_{ij})\in\mathbb{R}^6$.

Under (s1), we set $Q=6$, simulated $\log\phi_q\sim\text{N}(\log(\text{median}(\mathbf{D}_{\text{euc}})),2.25)$, computed $\boldsymbol{\Sigma}_q(\phi_q)$ from the dyadic covariance and the Matérn kernel of Section 4.1, and simulated $\mathbf{w}_q\sim\text{N}(\boldsymbol{0},\boldsymbol{\Sigma}_q(\phi_q))$ before drawing loadings from the global-local shrinkage prior. 
That prior drew each loading from a product of two half-Cauchy scales such that the loadings had no finite moments. Without such a step, the realized signal-to-noise ratio would vary significantly between replicates; rescaling $\boldsymbol{\Delta}$ to a fixed root mean square removed that variation. We chose a root mean square of 0.30 to have the same order as the root mean square of the posterior mean $\boldsymbol{\Delta}$ in the invaded-range fit, which was 0.38. Applying the same value under (s2) forced the misspecified process to have the same signal strength as the correctly specified one, which ensured that the comparison between them was not a comparison of signal-to-noise ratios.
Under (s2), $\boldsymbol{\Delta}$ was a deterministic step field over dyad midpoints; we partitioned the domain into quadrants and assigned each dyad the coefficient vector of the quadrant containing the midpoint of its two locations. As a result, the surface was piecewise constant, had no low-rank representation, was not a Gaussian process realization, and was discontinuous across the quadrant boundaries. Under (s3), $\boldsymbol{\Delta}=\boldsymbol{0}$ and the dyad errors followed a $t_5$ distribution rescaled to variance $\sigma^2$.

\subsection*{F.2 Recovery of the Scalar Parameters}
 
We report RMSE and coverage for $\boldsymbol{\beta}$ and for the coefficient surface in Table 1 of the manuscript. Supplemental Table \ref{tab:sim-scalars} reports the corresponding quantities for the scalar parameters.
 
\begin{table}[h]
\caption{Recovery of the scalar parameters in the replicated simulation study. Each cell is the mean over 25 replicates with the Monte Carlo standard error in parentheses; coverage is of nominal $95\%$ credible intervals.}
\label{tab:sim-scalars}
\centering\footnotesize
\setlength{\tabcolsep}{4pt}
\resizebox{\textwidth}{!}{%
\begin{tabular}{llcccccccc}
\toprule
 &  & \multicolumn{2}{c}{$\alpha$} & \multicolumn{2}{c}{$\sigma^2$} & \multicolumn{2}{c}{$\sigma^2_\eta$} & \multicolumn{2}{c}{$\phi_\eta$} \\
\cmidrule(lr){3-4}\cmidrule(lr){5-6}\cmidrule(lr){7-8}\cmidrule(lr){9-10}
Data Generating Process & Model & bias & cov. & bias & cov. & bias & cov. & bias & cov. \\
\midrule
\multirow{4}{*}{(s1): correct}
    & (i) standard & $\mathbf{-0.002}\,(0.014)$ & $0.88$ & $+0.936\,(0.048)$ & $0.00$ & $+0.162\,(0.022)$ & $0.04$ & $-0.028\,(0.007)$ & $1.00$ \\
    & (ii) + connectivity & $-0.005\,(0.013)$ & $0.92$ & $+0.791\,(0.054)$ & $0.00$ & $+0.154\,(0.022)$ & $0.04$ & $\mathbf{-0.023}\,(0.008)$ & $1.00$ \\
    & (iii) + DSVCs & $+0.007\,(0.011)$ & $0.92$ & $+0.111\,(0.015)$ & $0.16$ & $+0.038\,(0.004)$ & $0.32$ & $+0.043\,(0.009)$ & $1.00$ \\
    & (iv) full & $+0.004\,(0.006)$ & $1.00$ & $\mathbf{+0.014}\,(0.009)$ & $0.80$ & $\mathbf{+0.024}\,(0.002)$ & $0.60$ & $+0.067\,(0.008)$ & $1.00$ \\
\midrule
\multirow{4}{*}{(s2): misspecified $\boldsymbol{\Delta}$}
    & (i) standard & $+0.009\,(0.013)$ & $0.84$ & $+0.694\,(0.025)$ & $0.00$ & $+0.227\,(0.018)$ & $0.00$ & $-0.060\,(0.009)$ & $0.96$ \\
    & (ii) + connectivity & $+0.012\,(0.013)$ & $0.84$ & $+0.612\,(0.022)$ & $0.00$ & $+0.224\,(0.018)$ & $0.00$ & $-0.063\,(0.009)$ & $0.96$ \\
    & (iii) + DSVCs & $+0.021\,(0.010)$ & $0.72$ & $+0.125\,(0.014)$ & $0.12$ & $+0.034\,(0.004)$ & $0.36$ & $\mathbf{+0.045}\,(0.008)$ & $1.00$ \\
    & (iv) full & $\mathbf{+0.005}\,(0.004)$ & $1.00$ & $\mathbf{+0.001}\,(0.004)$ & $0.88$ & $\mathbf{+0.018}\,(0.001)$ & $0.76$ & $+0.084\,(0.009)$ & $1.00$ \\
\midrule
\multirow{4}{*}{(s3): no DSVCs, $t_5$ errors}
    & (i) standard & $\mathbf{+0.002}\,(0.005)$ & $1.00$ & $+0.068\,(0.003)$ & $0.00$ & $+0.025\,(0.002)$ & $0.40$ & $\mathbf{+0.048}\,(0.010)$ & $1.00$ \\
    & (ii) + connectivity & $+0.004\,(0.005)$ & $0.84$ & $\mathbf{+0.001}\,(0.003)$ & $0.76$ & $\mathbf{+0.015}\,(0.001)$ & $0.80$ & $+0.097\,(0.008)$ & $1.00$ \\
    & (iii) + DSVCs & $+0.007\,(0.005)$ & $0.96$ & $+0.034\,(0.005)$ & $0.28$ & $+0.023\,(0.002)$ & $0.48$ & $+0.060\,(0.010)$ & $1.00$ \\
    & (iv) full & $+0.004\,(0.005)$ & $0.92$ & $-0.008\,(0.003)$ & $0.64$ & $\mathbf{+0.015}\,(0.001)$ & $0.88$ & $+0.098\,(0.008)$ & $1.00$ \\
\bottomrule
\end{tabular}}
\end{table}
 
Every model estimated the intercept with a bias of at most 0.021 under every process, although coverage fell to 0.72 for the DSVC-only model under (s2). The two models
containing DSVCs recovered the dyadic variance, whereas the other two models did not. The bias in $\sigma^2$ under (s1) fell from $0.936$ for the standard model
to $0.014$ for the full model, and coverage rose from $0.00$ to $0.80$ because
a model with no coefficient surface absorbed the omitted surface into the
residual variance. The latent spatial variance was covered between $0.00$ and $0.88$, whereas the spatial
range parameter was over-covered for every model under every process.

\subsection*{F.3 Visual Model Comparison}

We provide a visual comparison of model performance in Supplemental Figure \ref{fig:sim-tiles}, where we reported predictive error and uncertainty for each dyadic response in the single simulated data set of our original submission, which was generated from the proposed model with $n=100$ nodes. 
In particular, we report the average absolute residual between the true dyadic response and the predicted response on the log1p scale: $\frac{1}{K}\sum^K_{k=1}\log(1+|\widetilde{y}_{ij}-y_{ij}^{(k)}|)$, where $y_{ij}^{(k)}$ denotes the predicted dyadic response between nodes $i$ and $j$ on the $k$th MCMC iteration. We used the log1p scale to compress the heavy right tail of absolute residuals while handling zero-valued residuals. 
To quantify uncertainty in model prediction, we used the posterior variability of the absolute residuals across MCMC samples. In Supplemental Figure \ref{fig:sim-tiles}, the color of the $(i,j)$ cell corresponds to the average absolute residual for the dyad comprising nodes $i$ and $j$ on the log1p scale, and the transparency of the cell corresponds to the uncertainty of the absolute residual; more transparent cells have more uncertainty, whereas more opaque cells have less uncertainty. Ideally, every cell has low average absolute residual and is opaque, indicating accurate and precise predictions. Models (i) and (ii) had higher average absolute residuals and less precise predictions, whereas models (iii) and (iv) had more accurate and precise predictions, as evidenced by the large number of opaque, blue cells. 
Therefore, models that accounted for connectivity and dyadic spatial variation yielded more accurate and precise predictions of processes exhibiting spatial heterogeneity (e.g., invasion dynamics).

\begin{figure}[H]
  \centering
  \begin{subfigure}{0.45\textwidth}
    \includegraphics[width=\linewidth]{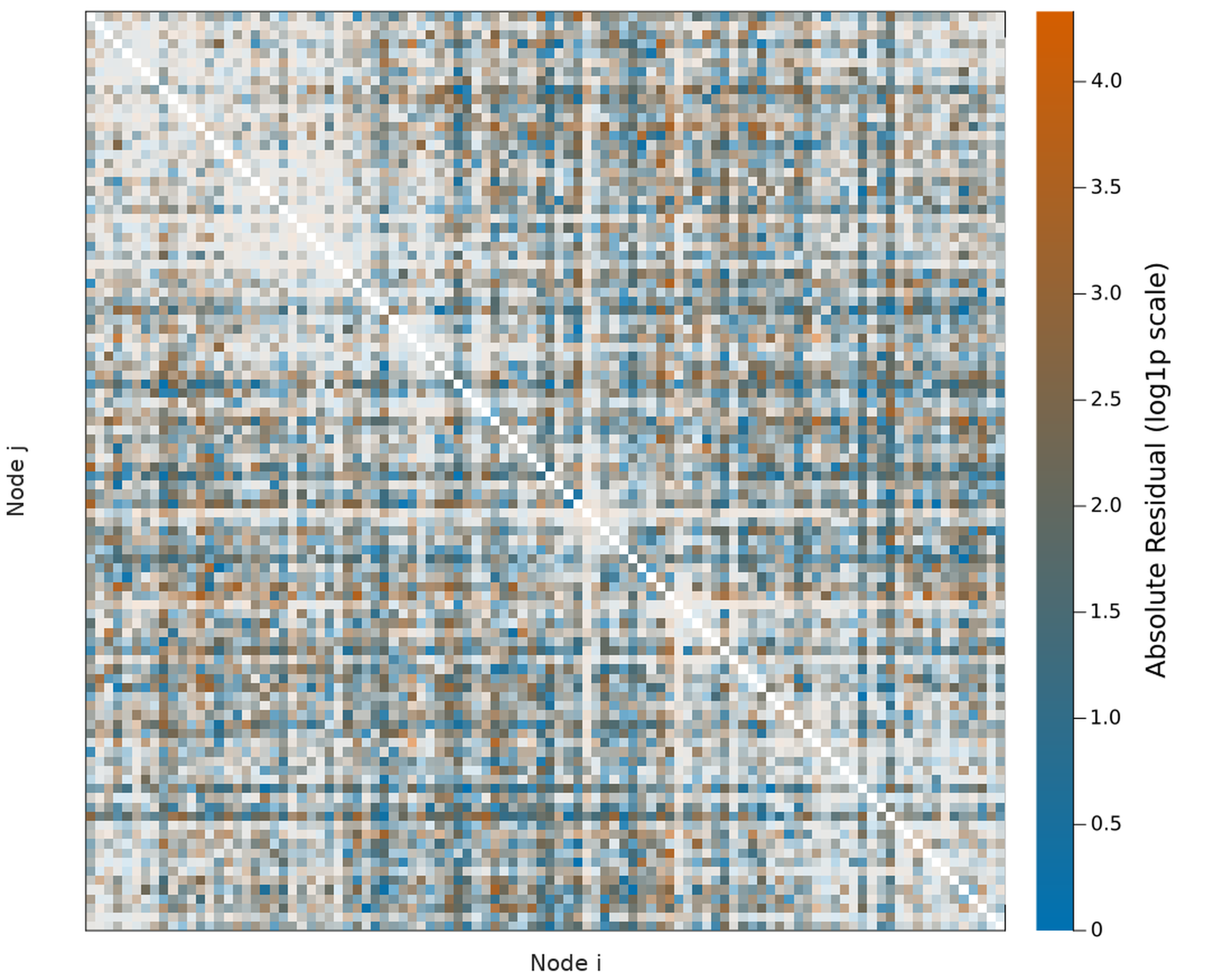}
    \caption{Model (i)}
    \label{fig:first}
  \end{subfigure}\hfill
  \begin{subfigure}{0.45\textwidth}
    \includegraphics[width=\linewidth]{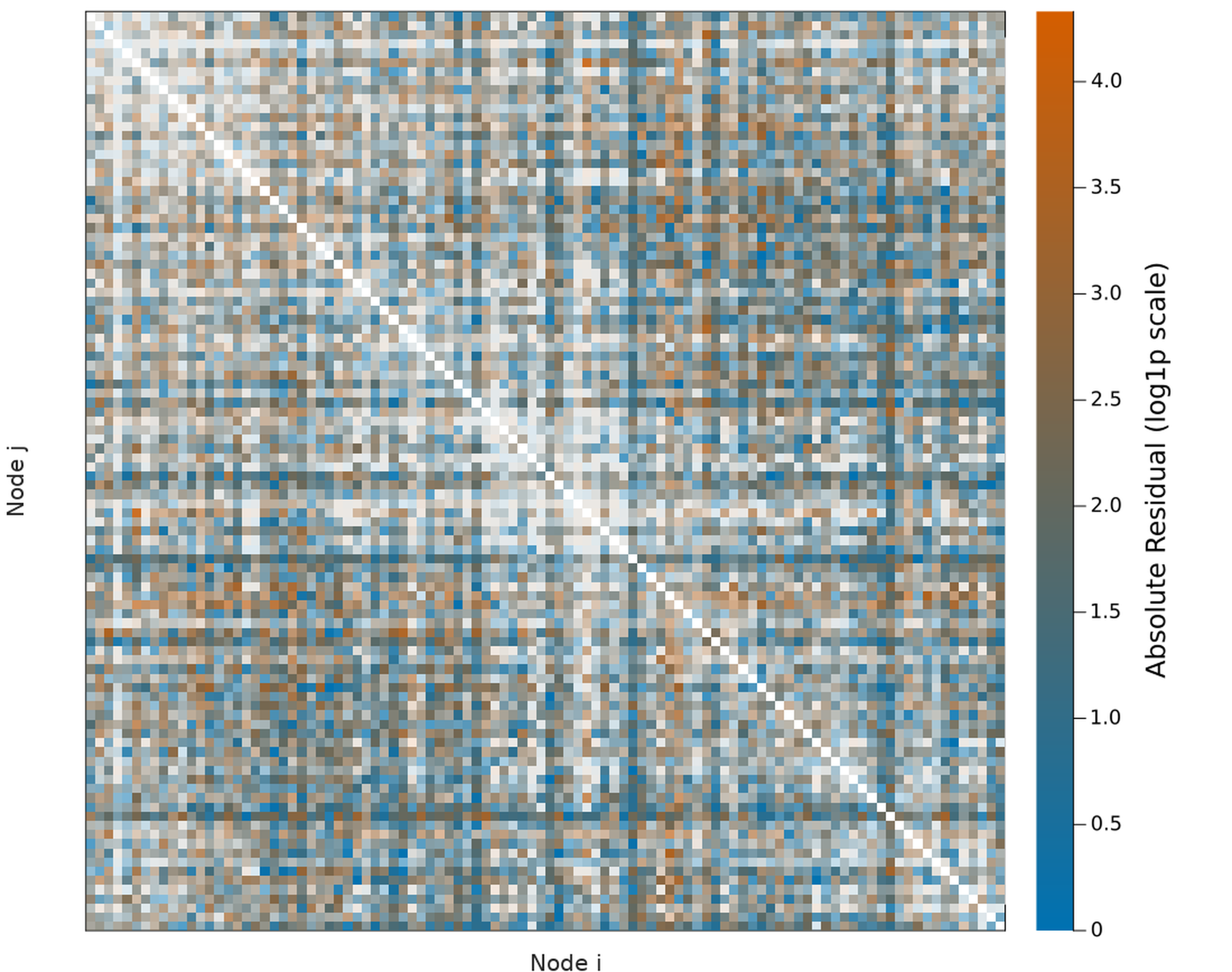}
    \caption{Model (ii)}
    \label{fig:second}
  \end{subfigure}

  \begin{subfigure}{0.45\textwidth}
    \includegraphics[width=\linewidth]{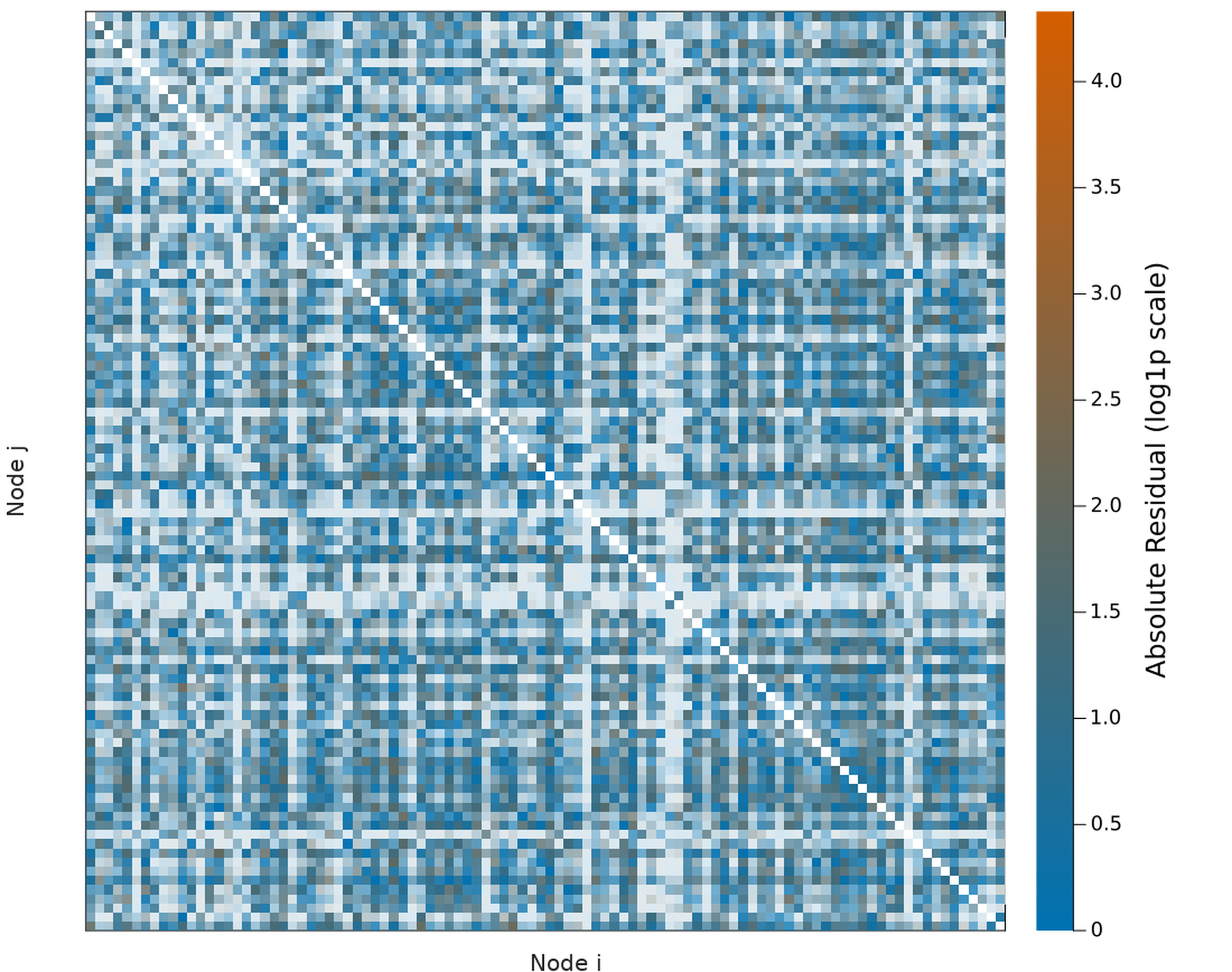}
    \caption{Model (iii)}
    \label{fig:third}
  \end{subfigure}\hfill
  \begin{subfigure}{0.45\textwidth}
    \includegraphics[width=\linewidth]{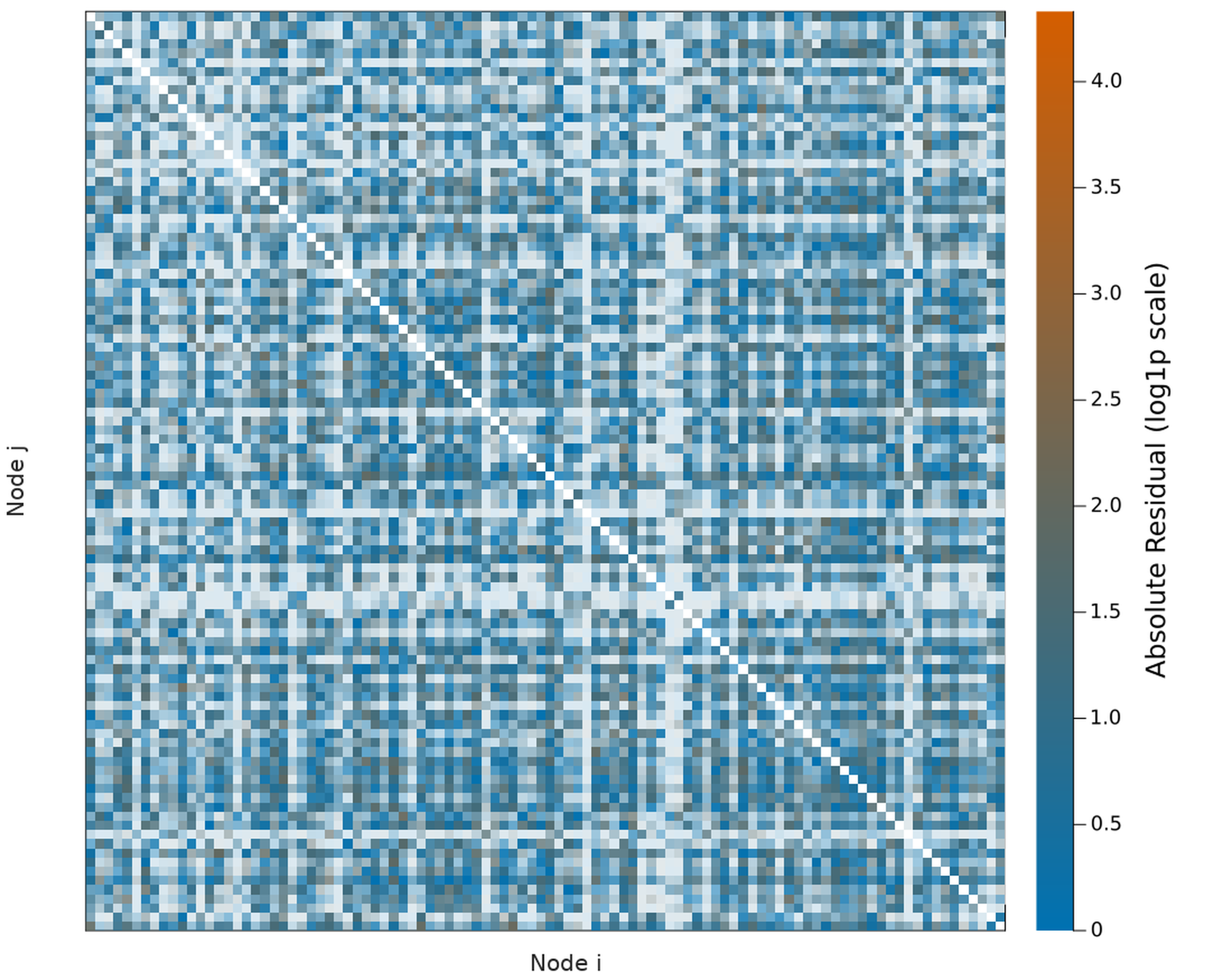}
    \caption{Model (iv)}
    \label{fig:fourth}
  \end{subfigure}

  \caption{The average absolute predictive error for each observed dyad with associated uncertainty for one representative replicate. Color represents the average absolute residual on the log1p scale, and transparency represents the uncertainty associated with each prediction. Dyads with more variable predictions have transparent cells, whereas dyads with more precise predictions have opaque cells.}
  \label{fig:sim-tiles}
\end{figure}

\section*{Appendix G - Supplemental Results for the Invaded Range}

\begin{figure}[H]
  \centering
    \includegraphics[width=\linewidth]{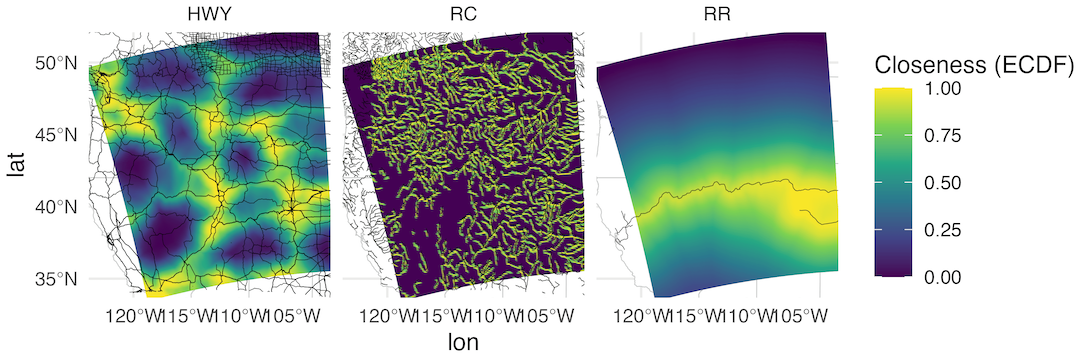}
    \label{fig:closeness}
  \caption{The node-level closeness measures $\mathbf{V}_c$ across the invaded range. The covariates ``HWY,'' ``RC,'' and ``RR'' denote the modern highways, rivers/streams, and historic railroads classes, respectively. For visualization, closeness scores are standardized and mapped using the class-specific empirical cumulative distribution function (ECDF) with colors truncated at the 99th percentile to suppress outliers.}
\end{figure}

\begin{figure}[h]
    \centering
    \includegraphics[width=0.7\linewidth]{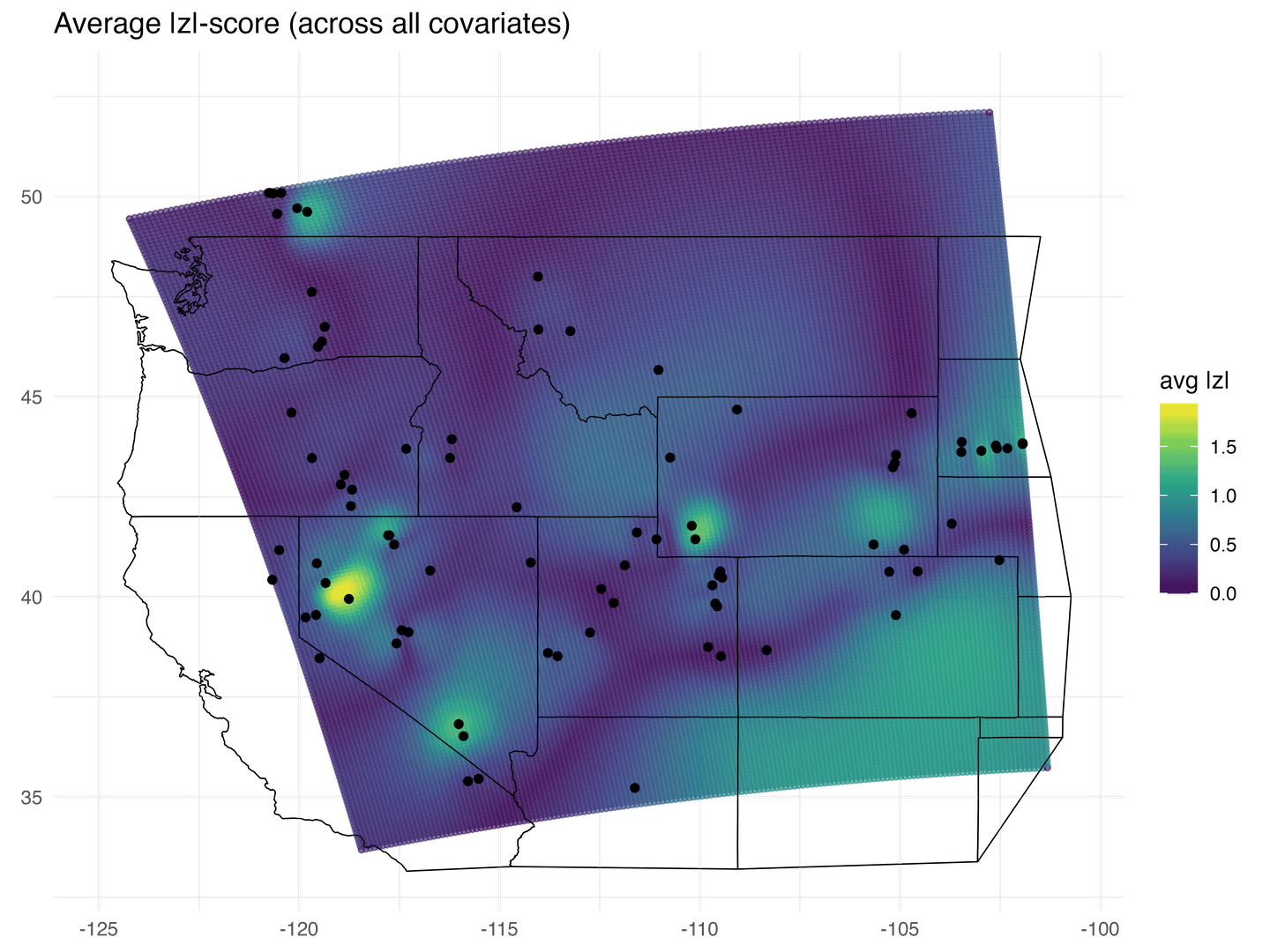}
    \caption{Average absolute $z$-scores of posterior predictive DSVCs, aggregated per grid cell across neighboring dyads and covariates. Higher average absolute $z$-scores indicate locations where spatially varying environmental effects were statistically significant. Observed locations are represented by black dots. See Section 7.2 for more details.}
    \label{fig:delta-z}
\end{figure}

\subsection*{G.1 Construction of the Mean Vector Field}

For each interior grid node $g\in\mathcal{G}$ with cardinal neighbors $\mathcal{N}(g)$, we estimated the expected genetic dissimilarity with neighbor $g'\in\mathcal{N}(g)$ as  
$$\mu_{(g,g')} = \alpha + \mathbf{z}_{ij}'(\boldsymbol{\beta} + \boldsymbol{\Delta}_{(g,g')}) 
+ (\eta_{g'} - \eta_g), $$
where $\mathbf{z}_{ij}=[\mathbf{h}(\mathbf{x}_{g'}-\mathbf{x}_g)' \;\boldsymbol\kappa'_{ij}]'$, $\mathbf{h}(\cdot)$ denotes the RBF transformation of pairwise differenced environmental effects, and  $\eta_g$ is the latent spatial random effect for node $g$. 
Then, we obtained directional gradients by finite differencing. For example, we computed the longitudinal and latitudinal components of the mean vector at node $g$ as  
$$u_g = \frac{\mu_{(g,E)} - \mu_{(g,W)}}{s_x(g)}, 
\qquad 
v_g = \frac{\mu_{(g,N)} - \mu_{(g,S)}}{s_y(g)}, $$ 
where $g'\in\{N,E,S,W\}$ are the cardinal neighbors of node $g$ and $s_x(g)$ and $s_y(g)$ denote the local grid spacing in longitude and latitude, respectively.  
The resulting set of vectors $\{(u_g,v_g):g\in\mathcal{G}\}$ defined a mean vector field that summarized the local magnitude and direction of gene flow across the landscape. The resulting field is in Figure 2 of the manuscript. 

\begin{figure}[h]
  \centering
    \includegraphics[width=\linewidth]{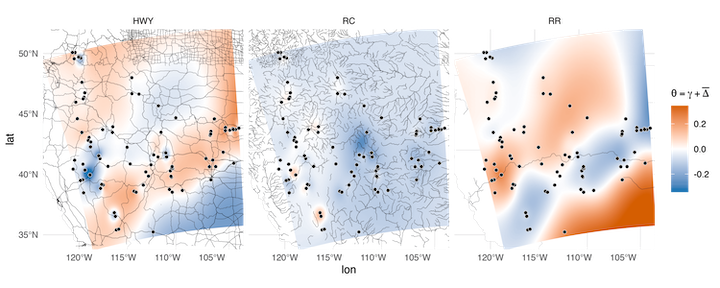}
    \label{fig:closeness}
  \caption{Posterior node-level slopes $\theta_{g,c}$ across the invaded range for the three connectivity covariates. Black lines depict the features in each class, and black dots denote observation locations. The spatial patterns resemble those of the average absolute $z$-scores of posterior predictive DSVCs (Supplemental Figure \ref{fig:delta-z}), which reveals localized, statistically significant regions  where the model infers nonstationary connectivity effects. Only 11 of the 86 sampled genotypes lie within 25km of either railroad feature. As a result, the railroad closeness measure has little contrast between dyads at any decay.}
\end{figure}

\subsection*{G.2 Predictive Residuals}

\begin{figure}[H]
  \centering
  \begin{subfigure}{0.45\textwidth}
    \includegraphics[width=\linewidth]{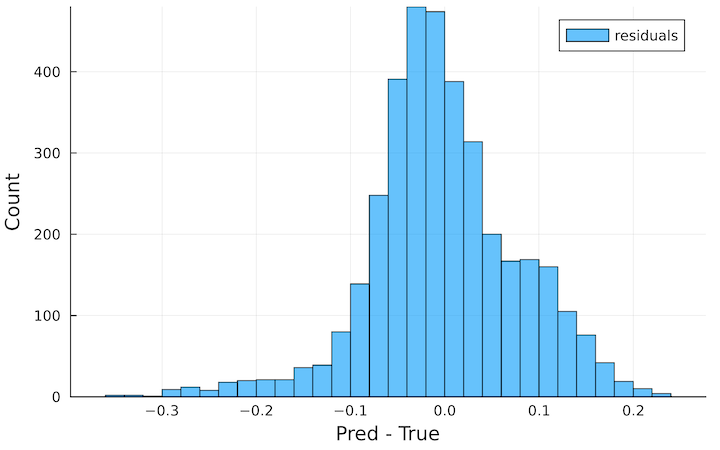}
    \caption{Kinship residuals under (i1).}
    \label{fig:first}
  \end{subfigure}\hfill
  \begin{subfigure}{0.45\textwidth}
    \includegraphics[width=\linewidth]{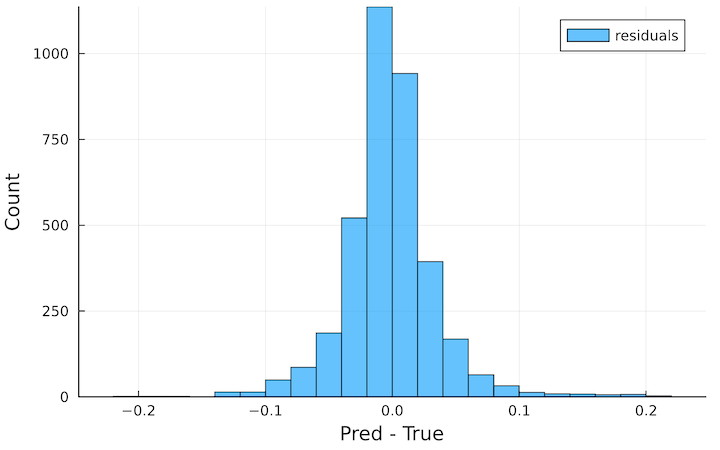}
    \caption{Kinship residuals under (i4).}
    \label{fig:second}
  \end{subfigure}

  \begin{subfigure}{0.45\textwidth}
    \includegraphics[width=\linewidth]{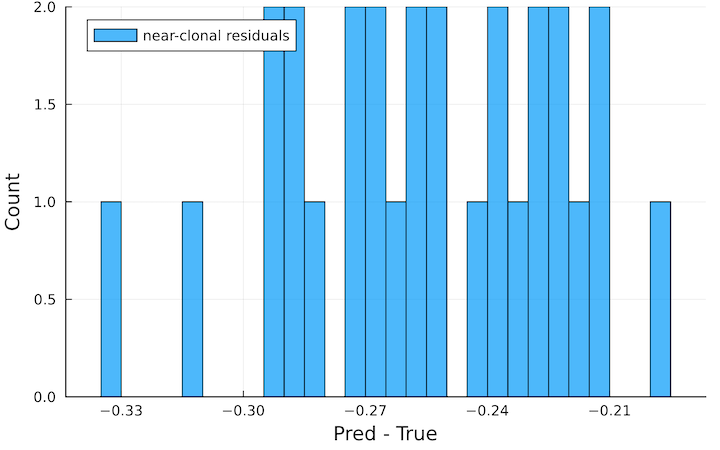}
    \caption{Near-clonal kinship residuals under (i1).}
    \label{fig:third}
  \end{subfigure}\hfill
  \begin{subfigure}{0.45\textwidth}
    \includegraphics[width=\linewidth]{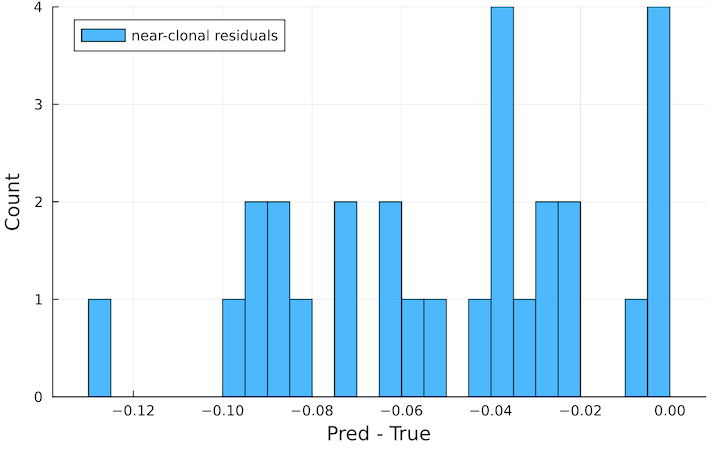}
    \caption{Near-clonal kinship residuals under (i4).}
    \label{fig:fourth}
  \end{subfigure}

  \caption{Posterior means of residuals between true kinship $1-\text{logistic}(\widetilde{y}_{ij})$ and predicted kinship $1-\text{logistic}(\widetilde{y}^*_{ij})$ for all pairs (a-b) and near-clonal pairs (c-d).}
  \label{fig:resid}
\end{figure}

In Section 7.2, we compare the kinship residuals under (i1) and (i4) across all dyads and on the near-clonal pairs; Supplemental Figure \ref{fig:resid} shows their distributions.

\section*{Appendix H - Computational Cost and Scaling}

\subsection*{H.1 Preprocessing}
 
Every step that precedes the first MCMC iteration in the invaded-range analysis
took 63.0 seconds, of which 5.6 seconds was package and library loading and 57.4
seconds was data preparation. Preparation includes loading the genetic distance matrix, which we computed once beforehand as described in Appendix A.1, computation of the closeness
measures and shared-segment scores across the three pathway classes comprising
767 features, and selection of the radial basis function centers. Compared to an MCMC algorithm that takes roughly 13 hours for the full model, preprocessing is
negligible and does not motivate any of the choices in Appendix A.
 
\subsection*{H.2 Scaling in the Number of Nodes}
 
The dominant per-iteration cost is the dense factorization of the $N\times N$
precision matrix described in Appendix E, which the sampler performs once for
each of the $Q$ latent factors. We measured the cost of the full model at
$n\in\{25,50,75,100,125,150\}$ over 300 iterations at each design size and report the results in Supplemental Table \ref{tab:h1}.
At $n=25$, an $N\times N$ matrix requires less
than a megabyte, per-iteration overhead costs more than the linear algebra, and a slope fitted across all six points gives an exponent of
$1.78$. However, the local exponents rise from $0.60$ to between $2.47$ and $2.63$ as the factorizations
begin to dominate.
 
\begin{table}[h]
\caption{Per-iteration cost and memory for the full model at various values of $n$. The local exponent is the slope between consecutive design sizes on the log scale in $N$ rather than in $n$. ``matrix'' is the size of a single $N\times N$ matrix in double precision, and ``RSS'' is the increase in memory over the baseline process.}
\label{tab:h1}
\centering\small
\begin{tabular}{rrrrrr}
\toprule
$n$ & $N$ & seconds per iteration & local exponent & matrix (GB) & RSS (GB) \\
\midrule
 25 &   300 & \phantom{0}0.027 & --- & 0.001 & \phantom{0}0.39 \\
 50 &  1225 & \phantom{0}0.062 & 0.60 & 0.011 & \phantom{0}0.56 \\
 75 &  2775 & \phantom{0}0.365 & 2.16 & 0.057 & \phantom{0}1.43 \\
100 &  4950 & \phantom{0}1.676 & 2.63 & 0.183 & \phantom{0}3.00 \\
125 &  7750 & \phantom{0}5.329 & 2.58 & 0.448 & \phantom{0}7.17 \\
150 & 11175 & 13.159 & 2.47 & 0.930 & 12.72 \\
\bottomrule
\end{tabular}
\end{table}
 
\subsection*{H.3 Memory}
 
Memory (rather than time) is the bottleneck in our analyses (see Supplemental Table \ref{tab:h1}). Additionally, the sampler holds $Q$ Cholesky factors alongside the current
factor scores. Memory reached 12.7 GB at $n=150$ on the
64 GB machine used for our analyses; the quadratic growth indicates that the limit for such a machine is approximately $n=200$ individuals.

\end{document}